\documentclass[aip,pop,amsmath,amssymb,reprint]{revtex4-1}%
\usepackage[dvipdfmx]{graphicx}
\usepackage{dcolumn,bm}
\usepackage[english]{babel}
\usepackage{float}
\usepackage[T1]{fontenc}
\usepackage[dvips,colorlinks,linkcolor=blue,citecolor=blue,urlcolor=blue]{hyperref}
\hypersetup{pdftitle={Instabilities of collisionless current sheets revisited: the role of anisotropic heating},pdfauthor={P. A. Mu\~noz}}

\begin{document}
\title[]{
	Instabilities of collisionless current sheets revisited: the role of anisotropic heating
}

\author{P. A. Mu\~noz}
\email{munozp@mps.mpg.de}
\author{P. Kilian}
\email{kilian@mps.mpg.de}
\author{J. B\"uchner}
\email{buechner@mps.mpg.de}
\affiliation{Max-Planck-Institut f\"ur Sonnensystemforschung, 37077 G\"ottingen, Germany}

\date{\today}

\begin{abstract}
	In this work, we investigate the influence of the anisotropic heating on the spontaneous instability and evolution of thin Harris-type collisionless current sheets, embedded in antiparallel magnetic fields. In particular, we explore the influence of the macroparticle shape-function using a 2D version of the PIC code ACRONYM. We also investigate the role of the numerical collisionality due to the finite number of macroparticles in PIC codes. It is shown that it is appropriate to choose higher order shape functions of the macroparticles compared to  a larger number of macroparticles per cell. This allows to estimate better the anisotropic electron heating due to the collisions of macroparticles in a PIC code. Temperature anisotropies can stabilize the tearing mode instability and trigger additional current sheet instabilities. We found a good agreement between the analytically derived threshold for the stabilization of the anisotropic tearing mode and other instabilities, either spontaneously developing or initially triggered ones. Numerical effects causing anisotropic heating at electron time scales, become especially important for higher mass ratios (above $m_i/m_e = 180$). If numerical effects are carefully taken into account, one can recover the theoretical estimated linear growth rates of the tearing instability of thin isotropic collisionless current sheets, also for higher mass ratios.\\

	\textit{Copyright (2014) American Institute of Physics. This article may be downloaded for personal use only. Any other use requires prior permission of the author and the American Institute of Physics.\\
The following article appeared in P. A. Mu\~noz, P. Kilian, and J. B\"uchner, Physics of Plasmas \textbf{21}, 112106 (2014),   and may be found at
\href{http://dx.doi.org/10.1063/1.4901033}{http://dx.doi.org/10.1063/1.4901033}
 }
\end{abstract}

\maketitle
\section{Introduction}\label{sec:intro}
Current sheets  (CS) are ubiquitous in space plasmas, as those of the solar corona or of the Earth's magnetotail. They can become unstable and enable the dissipation of magnetic energy by reconnection, causing turbulence, plasma and particle acceleration\cite{Buchner2007}. In fact, CS are the preferred locations where magnetic reconnection can take place. CS are prone to a number of different instabilities driven by currents gradients or streams as source of free energy\cite{Silin2002,Silin2003}. We focus on thin CS, that have a halfwidth of the order of the ion inertial length. This is supposed to be the natural limit of their thinning by compression. Inside of the ion inertial length, particles may experience resonant wave-particle interactions which may provide the physical mechanism for dissipation in collisionless plasmas~\cite{Buchner2007a,Galeev1984}. In fact, many space plasmas in the solar system and beyond are collisionless, hence thin CS are of fundamental importance in those environments, e.g., for the onset of solar flares and substorms\cite{Runov2003}.

The most fundamental instability of CS is the tearing instability which produces magnetic islands, i.e.: a change of topology necessary for magnetic reconnection. It was first investigated for collision-dominated plasma in the framework of Magnetohydrodynamics (MHD) approach\cite{Furth1963}. Later, it was proposed and investigated also for collisionless plasmas\cite{Coppi1966,Schindler1974,Drake1977,Galeev1979}. The mechanism for the collisionless tearing instability and for its non-linear saturation depends on the specific parameters of the CS. It can be the lack of gyrotropy of the electron distribution function, the electron pressure anisotropy and/or the electron trapping (see Ref.~\onlinecite{Treumann2013b} for a recent review).

Previous studies have shown that the collisionless tearing instability is very sensitive to electron temperature anisotropy\cite{Chen1984,Daughton2004,Haijima2008,Quest2010a}. On the other hand, it is known that numerical simulations by Particle-in-Cell (PIC)\cite{Buchner1997} and Vlasov codes \cite{Buchner2006a}  may produce artificial heating under some circumstances (see Refs.~\onlinecite{Birdsall1991,Hockney1988}  and references therein).
Numerical heating may mimic physical heating processes and, like those, it may affect the growth rate of the tearing instability. In addition, it can cause instabilities that require pre-heating for their development. Two independent numerical heating mechanisms were identified (see Ref.~\onlinecite{Cormier-Michel2008} and references therein): grid heating and scattering. According to Ref.~\onlinecite{Cormier-Michel2008}, grid heating is due to a
kinetic instability resulting from the aliasing of high frequency modes not resolved by the grid. Hence, grid heating is important when the Debye length is not resolved well enough by the grid. Since one can avoid it by resolving the Debye length, we will not analyze or discuss this well-known process here. Instead, we focus on the numerical enhanced particle heating due to scattering. This unphysical process arises due to a finite number of macroparticles per cell replacing, as an approximation, the continuous phase space. When a macroparticle moves to the next cell, it produces locally random fluctuations of the electric potential that will act backwards on the other macroparticles. Thus, macroparticles experience an effective scattering due to these non-physical forces which are absent in real collisionless plasmas (see, e.g., Refs.~\onlinecite{Langdon1970,Abe1975}). Therefore, this scattering can be interpreted as a form of numerical collisions among the macroparticles.

While in the past extensive studies of the  grid scale-induced numerical heating and their effects in PIC simulations were carried out, the heating due to numerically caused macroparticle scattering remained less well analyzed. Just recently, Cormier-Michel et al.\cite{Cormier-Michel2008} showed that there are some effects in  laser wakefield accelerators depending critically on some threshold, above which numerical scattering dominates those phenomena. Furthermore, these authors found that the
choice of an appropriate shape function of macroparticles can be much more efficient in reducing the numerical scattering and heating than the usual approach of increasing the number of macroparticles per cell. See Appendix~\ref{sec:shape_function} for the definition and importance of the shape function (a form of interpolation scheme) in PIC codes. For a discussion about the dependence of the numerical scattering/collisions on the shape function, see Appendix~\ref{sec:collisions}.

The relation between shape functions and heating can be understood in the following way: the stochastic force producing the numerical scattering is introduced by errors in the interpolation of the electromagnetic field at the macroparticles position. Thus, it leads to errors in the resulting momenta of the macroparticles, such as their temperature~\cite{Cormier-Michel2008}.  Therefore, the numerical heating will be reduced by smoother interpolations, since they will reduce the non-physical forces by smoothing out the sharp border of the macroparticles (see Sec.~13.5 of Ref.~\onlinecite{Birdsall1991}).

So far, the influence of the use of higher order shape functions in PIC simulations on the stability of collisionless Harris CS has not been explicitly investigated yet. For this reason, the main purpose of this paper is to study the influence of the shape functions on heating processes in the course of the CS evolution, as well as to distinguish their importance compared to other numerical processes. As we will show, if those numerical considerations are not taken into account, the simulations may lead to non-physical results.

This paper is organized as follows. In Sec.~\ref{sec:setup}, we describe the simulation setup. In Sec.~\ref{sec:heating}, we discuss the consequences of the numerically-induced anisotropic heating.  In Sec.~\ref{sec:anisotropic_runs}, we analyze the origins and consequences of an initially imposed temperature anisotropy. In Sec.~\ref{sec:tearing}, we analyze the influence of these numerical effects on the development of the tearing instability in a Harris CS. Finally, we summarize our findings in the conclusion, Sec.~\ref{sec:conclusion}.
\section{Simulation setup}\label{sec:setup}
We consider the stability of CS in a Harris equilibrium~\cite{Harris1962}, with a magnetic field changing as
\begin{equation}\label{harris}
	\vec{B}(x) = B_{\infty y}\tanh\left(x/L\right)\hat{y},
\end{equation}
where $L$ is the halfwidth of the CS and $B_{\infty y}$ is the asymptotic magnetic field. The Harris equilibrium is sustained by the current carried by counterstreaming electrons and ions (denoted with the subindices $e$ and $i$) with drifting speeds $U_{Di}$ and $U_{De}$, respectively. This current density is given by:
\begin{equation}\label{harris-current}
	\vec{J}(x) = 2en_0U_{Di}\cosh^{-2}\left(x/L\right)\hat{z} \qquad\text{and}\qquad \frac{U_{Di}}{T_i} = - \frac{U_{De}}{T_e}.
\end{equation}
The latter condition guarantees the absence of initial electric fields. The asymptotic magnetic field and the peak density at the center of the CS (with $n_0=n_{0e}=n_{0i}$) are related by the equilibrium condition between the magnetic and thermal pressures
\begin{equation}\label{harris_equilibrium}
	\frac{|B_{\infty y}|^2}{2\mu_0}=n_{0}(k_B T_e + k_B T_i).
\end{equation}
We do not impose a background density in the first place, in order to minimize additional effects due to that population and in that way to reduce the number of free parameters to be chosen. Note that we will discuss the consequences of a background plasma in Sec.~\ref{sec:background}

On purpose, we do not impose an initial perturbation other than the shot noise due to the finite number of macroparticles, because we focus on the onset and growth of the spontaneous instabilities\cite{Daughton2005,Pritchett2005}. Hence, the system evolves only from this macroparticle noise. Note that simulations with an initial perturbation bypass this stage allowing the system to reach directly the phase of fully developed fast reconnection\cite{Birn2001}.

\begin{figure}[H]
	\centering
	\includegraphics[width=0.6\linewidth]{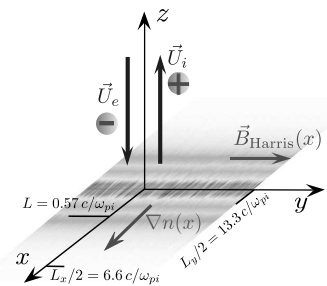}
	\caption{Schematics of the geometry of the simulation setup. The magnetic field $\vec{B}$ is in the $y$ direction on dependence on the $x$ coordinate. Correspondingly, the density $n$ and current density $\vec{J}$ vary along the $x$ direction. The $2D$ simulations are carried out in the $x-y$ plane. The CS is sustained by counterstreaming electrons and ions that produce a current density $\vec{J}$ in the out-of-plane direction $z$.} \label{fig:simulation_setup}
\end{figure}

See Fig.~\ref{fig:simulation_setup} for a schematics of the setup.
The four parameters necessary to fully specify this Harris sheet are chosen according to Ref.~\onlinecite{Lee2012f}:
\begin{equation}
	\frac{L}{c/\omega_{pi}}=0.57 ,\quad \frac{\omega_{pe}}{\Omega_{ce}}=2.87,\quad \frac{m_i}{m_e}=180,\quad \frac{T_i}{T_e}=1,
\end{equation}
where $c$ is the speed of light, $\omega_{pe}=\sqrt{n_0 e^2/(\epsilon_0 m_e)}$ is the electron plasma frequency (calculated with the central density  $n_0$), $\Omega_{ce}=eB_{\infty y}/m_e$ is the electron cyclotron frequency in the asymptotic  magnetic field.
The chosen mass ratio is a compromise between computational speed and the possibility to separate the effects of ion and electron motion. The width of the CS is of the order of the ion inertial length which is supposed to be the natural limit of the thinning of CS. The ratio of frequencies $\omega_{pe}/\Omega_{ce}$ is chosen to be relatively small to save computation time. Although this parameter may affect the nonlinear saturation of the tearing mode, the linear growth rate was shown to be independent of it\cite{Daughton2005}.

The choice of equal temperatures for ions and electrons implies that their drift speeds have the same magnitude but opposite directions: $\vec{U}_{Di}=-\vec{U}_{De}= U_D\hat{z}$. This also implies~\cite{Coroniti1977a} that the electron contribution dominates the growth of the tearing mode instability. The frequency ratio  $\omega_{pe}/\Omega_{ce}$ and the halfwidth $L$ are related with the electron thermal speed  $v_{th,e}=\sqrt{k_BT_e/m_e}$ and the initial drift speed, respectively, by means of
\begin{equation}
	\frac{\omega_{pe}}{\Omega_{ce}}=\frac{1}{2}\frac{c}{v_{th,e}}\qquad  \text{and}\qquad   \frac{\rho_i}{L}=\frac{1}{2}\frac{U_D}{v_{th,i}},
\end{equation}
with $\rho_i=v_{th,i}/\Omega_{ci}$ being the thermal ion gyroradius in the asymptotic magnetic field $B_{\infty y}$. In the results to be shown, lengths are normalized to the ion inertial length $d_i=c/\omega_{pi}$ and times to $\Omega_{ci}^{-1}$.

For the simulations, we use the  PIC code ACRONYM (see Ref.~\onlinecite{Kilian2012} for a description), which allows to select different shape functions, including all those described in the Appendix \ref{sec:shape_function}.  We utilize the code in its 2D3V mode, i.e.: no variations are calculated in the translational invariant $\hat{z}$ direction parallel to the current.
The boundary conditions for particles and electromagnetic fields are reflecting in the $\pm x$ direction and periodic in the $\pm y$ direction.

Other numerical parameters used in the simulations are the following.  The grid size $\Delta x$ is one Debye length $\lambda_{De}=\sqrt{\epsilon_0 k_B T_e/(n_0 e^2)}$ (in terms of our length unit, $(c/\omega_{pi})/\Delta x=76.8$). The time step $\Delta t=0.087\omega_{pe}^{-1}$ is chosen small enough to solve the electron motion and to fulfill the Courant condition $\Delta x/(c\Delta t)=0.5 < 1$. Following Ref.~\onlinecite{Lee2012f}, the simulation box is $L_x \times L_y=$ [1048 $\lambda_{De}$] x [2048 $\lambda_{De}$]=[13.3 $c/\omega_{pi}$] x [26.6 $c/\omega_{pi}$], i.e.: the number of grid points is 1048 x 2048. The total number of macroparticles per species is $7.28\cdot10^6$. This corresponds to 40 macroparticles per cell in the center of the CS.

The size in the $y$ direction was chosen to allow, according to the linear theory~\cite{Coppi1966},  the development of up to seven unstable tearing modes with wavelength $\lambda=L_y/M$, with $M$ an integer satisfying  $2\pi M L/L_y =k_y L \leq 1$ (for our parameters, $2\pi L/L_y=0.136$). Thus, we can investigate the interaction and exchange of energy between magnetic islands of different size, i.e.: multimode tearing. Note that for a simulation box size that would allow only one tearing island, this would reach stabilization and saturation stage at very small amplitudes, unimportant for space plasmas\cite{Karimabadi2005a}.
\section{Origin and consequences of the numerically generated anisotropic heating}\label{sec:heating}
\subsection{Numerical heating and shape function}\label{sec:heating_shapefunction}
In this section, we show the results of five simulation runs with different combinations of shape functions (CIC or TSC schemes) and macroparticle number (ppc). All the other parameters are the same. They are denoted as CIC-40ppc, CIC-160ppc, CIC-360ppc, TSC-40ppc and TSC-160ppc. The run CIC-40ppc uses 40 macroparticles per cell in the center of the CS, as previously described, and the other CIC-160ppc and CIC-360ppc use 4 and 9 times more particles, respectively.

The ACRONYM code is based on the momentum-preserving scheme for PIC codes. Hence, due to the discretization error and the distribution of macroparticles according to the shape function, the total energy $E_t$ is almost but not completely conserved~\cite{Abe1975}. As mentioned before, the small deviation is due to the unphysical non-conservative forces experienced by the macroparticles (see 7.6 of Ref.~\onlinecite{Hockney1988}). For this reason, our first check is the energy conservation. In PIC codes, this quantity is computed as follows,
\begin{eqnarray}
	E_t=&&\int\frac{B^2}{2\mu_0}\,d^3\vec{x}+ \int\frac{\varepsilon_0 E^2}{2}\,d^3\vec{x} \\
	 && + \sum_s\left(\frac{m_sn_s}{2}\vec{V}_s^2  + \frac{m_s}{2}\int (\vec{v}-\vec{V}_s)^2 f_s\, d\vec{v}^3\right).\nonumber
\end{eqnarray}
The terms in the right hand side are the magnetic, electric, kinetic and thermal energy, respectively. The sum is over all the species $s$. $\vec{V}_s=(1/n_s)\int \vec{v} f_s\,d\vec{v}^3$  is the bulk velocity, $n_s=\int f_s\,d\vec{v}^3$ the number density, $f_s$ the distribution function and $m_s$ the mass of each species.

\begin{figure}[H]
	\centering
	\includegraphics[width=\linewidth]{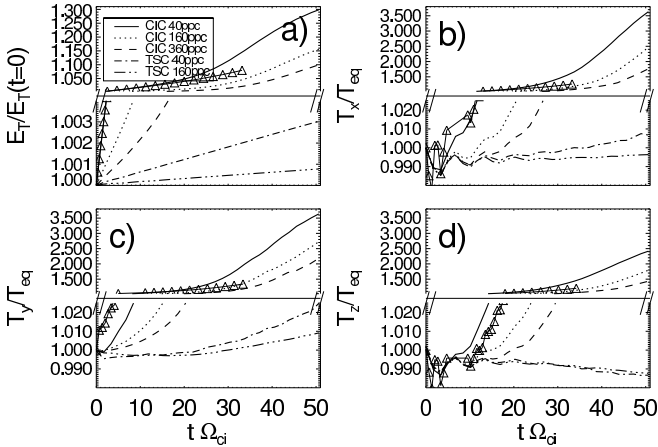}
	\caption{Time history of some quantities for the runs with different combinations of shape functions (CIC and TSC) and number of macroparticles per cell (40-360 ppc). It also includes a run with the XOOPIC code (using CIC shape function and 40ppc) for validation purposes (denoted by $\triangle$). The plots show: a) total energy (normalized to its initial value), b) average electron temperature $T_{e,x}$, c) average electron temperature $T_{e,y}$ and d) average electron temperature $T_{e,z}$. The temperatures are normalized to their equilibrium (initial) values. Note that we break the plots in the vertical direction in two parts, with larger scales in the upper parts. This is in order to visualize more easily the small variations of the quantities obtained using the TSC shape function, compared with the large variations in those obtained using the CIC shape function.}\label{fig:0b_temp_comparison}
\end{figure}

The results of our simulation runs for the evolution of the total energy are depicted in Fig.~\ref{fig:0b_temp_comparison}(a).
It can be seen that beyond $t\gtrsim 40\,\Omega_{ci}^{-1}$, the conservation of energy is improved by two orders of magnitude using the TSC over the CIC shape functions for the cases with 40 macroparticles per cell: the total energy for TSC-40ppc increases above its initial value by about $0.25\%$, while for CIC-40ppc by about $25\%$. A similar effect is seen for the case with 160 macroparticles per cell: the total energy for TSC-160ppc increases above its initial value by about $0.05\%$, while for CIC-160ppc by about $8\%$. Fig.~\ref{fig:0b_temp_comparison}(a) also demonstrates that an increase of the number of macroparticles does not have such strong effect, while slows down the simulation to a large extent. In fact, since the noise level scales as the square root of the number of macroparticles\cite{Birdsall1991}, a reduction of numerical noise to $50\%$ requires four times the number of macroparticles and thus it is four times slower (see e.g, CIC-160ppc compared with CIC-40ppc). However, the choice of TSC over CIC costs only a fraction of that effort and it slows the run by only $\sim 32\%$.
The apparent bad energy conservation of the runs with the CIC scheme deserves an additional validation test. For this sake, we also compare with the results obtained by using the XOOPIC code (Ref.~\onlinecite{Verboncoeur1995}), that only provides linear weighting (CIC). The parameters of this run are identical to the ACRONYM run CIC-40ppc. One can see in Fig.~\ref{fig:0b_temp_comparison}(a), the total energy evolution obtained by ACRONYM and XOOPIC. They are similar in the beginning, and start to diverge after $t\Omega_{ci}\sim16$. From that time onwards, XOOPIC has a better performance (the energy is conserved in $\lesssim 8\%$ for the last time depicted $t\Omega_{ci}\sim33$) than ACRONYM if the CIC scheme is used (energy conservation $\lesssim 16\%$ for $t\Omega_{ci}\sim33$). We checked that all the physical and numerical processes to be discussed in the next sections  develop almost identically for either of the two codes, at least until the aforementioned time. This allows us to conclude that the ACRONYM runs with the CIC shape function predict a similar behaviour of the modeled system compared to other standard PIC codes. Note that we have not obtained results with the XOOPIC code for times greater than $t\Omega_{ci}\sim33$, since we were using the free serial version of the code, which is much slower than ACRONYM.

Although not shown here, we also checked that the use of even higher order shape functions, such as PQS, does not improve significantly the conservation of energy in comparison with TSC (while slowing down the runs by $\sim 71\%$ compared to the CIC scheme). This is in agreement with another recent study (Ref.~\onlinecite{Cormier-Michel2008}). Those authors found that the influence of nonphysical processes due to the choice of the shape functions are much less notorious comparing cubic and quadratic interpolations rather than quadratic and linear ones.

We also found that the insufficient conservation of energy in the simulations using CIC shape functions is mainly due to artificial electron heating. As expected due to their larger inertia, the numerical ion heating is negligible. This can be seen in Fig.~\ref{fig:0b_temp_comparison}(b), depicting the evolution of components of the electron temperature $T_{e,x}$ averaged over the whole simulation box. The evolution of the electron temperature in the other two directions is displayed in Figs.~\ref{fig:0b_temp_comparison}(c) and \ref{fig:0b_temp_comparison}(d). All those curves follow the same trend as the total energy: the TSC-runs keep the temperatures more constant than the CIC-runs, while an increase in the macroparticle number does not help to the same extent. Hence, we notice that a higher order shape function allows to resolve better the details of the electron motion than a simple brute force increase of the macroparticle number.

It is important to note that this numerical electron heating is anisotropic: the increase in the out-of-plane temperature $T_{e,z}$ is less notorious than $T_{e,x}$ or $T_{e,y}$ (cf. Figs.~\ref{fig:0b_temp_comparison}(b)-\ref{fig:0b_temp_comparison}(d)). We will explore the origin of this behaviour in the next subsections.

Therefore, we showed that the choice of a higher order shape functions leads to a better conservation of energy and reduction of numerical heating in a more efficient and computationally less expensive way than the usually used increase of number of macroparticles per cell. This reduction of numerical heating is in agreement with the findings of Cormier-Michel~\cite{Cormier-Michel2008} for simulations of laser wakefield accelerators. Note that the role of shape functions was already discussed in early electrostatic PIC simulations comparing NGP and CIC schemes (see Ref.~\onlinecite{Hockney1971}).
\subsection{Suppression of tearing instability by bifurcation due to anisotropic heating: ''initially  isotropic runs``.}
\begin{figure}[H]
	\centering
	\includegraphics[width=\linewidth]{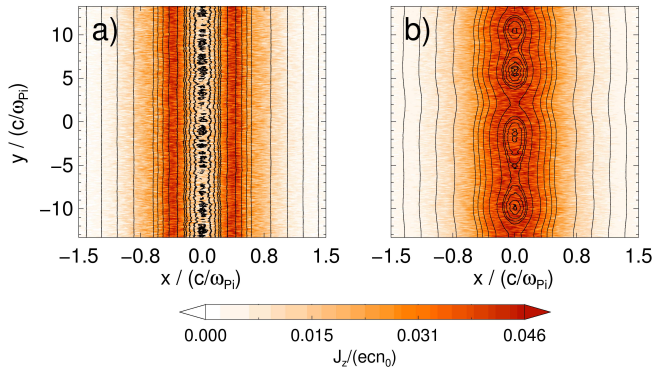}
	\caption{Contours of the out-of-plane component $J_z$ of the total current density, at the time $t=33\,\Omega_{ci} ^{-1}$, for two runs a) CIC-40ppc and b) TSC-40ppc. Black lines are the magnetic field lines. The run CIC-40ppc shows a bifurcated structure while TSC-40ppc shows only a single peak structure with large tearing mode islands.}\label{fig:bifurcation}
\end{figure}

\begin{figure}[H]
	\centering
	\includegraphics[width=\linewidth]{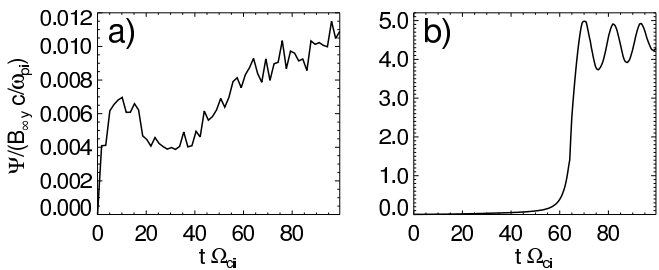}
	\caption{Comparison of reconnected flux $\Psi$ for the runs a) CIC-40ppc and b) TSC-40ppc. The left panel a) (with CIC shape function) correspond to the case with anisotropic heating and CS bifurcation, while the right panel b) (with TSC shape function) correspond to a tearing instability without significant anisotropic heating nor bifurcation.}\label{fig:m180_shape_fluxes}
\end{figure}

\begin{figure}[H]
	\centering
	\includegraphics[width=0.70\linewidth]{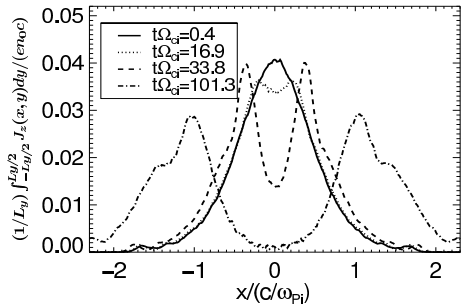}
	\caption{Evolution of the total current density profile $J_z$ showing bifurcation (CIC-40ppc run). The profiles are obtained by integrating the current density along the CS: $J_z(x) = (1/L_y) \int_{-Ly/2}^{Ly/2} J_z(x,y) dy$. }\label{fig:m180_cic_0b_cb256-multiple-Jz-profile_x-zs0}
\end{figure}

Here we discuss the physical consequences on the CS evolution of the anisotropic numerical heating seen in the runs with CIC shape function and/or an insufficiently small number of macroparticles per cell. Previous studies  have shown that a temperature anisotropy can generate bifurcated CS~\cite{Matsui2008a,Sitnov2003,Zelenyi2004,Daughton2004}. Temperature
anisotropies, generated e.g. in simulations with CIC shape functions, can produce bifurcation in the out-of-plane component of the total current density $J_z$\cite{Lee2012f}. A comparison for a specific time between the runs CIC-40ppc and TSC-40ppc is shown in Fig.~\ref{fig:bifurcation}. $J_z$ shows a double peak structure in the first run but not in the second one. Instead of that, TSC-40ppc shows several magnetic islands. This represents a growth of the tearing mode.
The different behaviour of the evolution of CS with CIC and TSC shape functions can also be shown by means of the evolution of reconnected flux $\Psi=\int \vec{B}\cdot\,d\vec{S}$. In a 2D setup, this quantity can be obtained as the difference of the vector potentials $A_z(x=0)$ between the O and X points of each magnetic island. Since the number of magnetic islands varies with time, and at the end there is only one left, we chose to compute this quantity by taking the difference between the maximum and minimum value of $A_z$ along the line $x=0$. This is a representative quantity when the islands can be easily distinguished, although not accurate at the beginning when there are many small magnetic islands arising from numerical noise. The evolution of the reconnected flux $\Psi$ for those cases is displayed in Fig.~\ref{fig:m180_shape_fluxes}.
The fast growth of the reconnected flux in Fig.~\ref{fig:m180_shape_fluxes}(b) after $t\sim60\,\Omega_{ci}^{-1}$ is the signature of the explosive phase of reconnection, after which there is only one remaining magnetic island in the whole simulation box. Afterwards, a saturated stage is reached where the entire structure of the CS is disrupted due to the outflows in opposite directions. Note that the latter is a numerical artifact due to the limited size of the simulation box  and the periodic boundary conditions in the $y$ direction. The values of the maximum reconnected flux $\Psi$ are of the same order of magnitude as reported in previous studies (see, e.g., Ref.~\onlinecite{Pritchett2005}).
Note that the reconnected flux in simulations with CIC shape function does not increase significantly. Instead, it always remains around the same value even for very long times, no magnetic islands are formed at the center of the CS, and therefore there are no X nor O points. This issue will discussed in more detail in Sec.~\ref{sec:tearing}.
In order to see more clearly the evolution of this bifurcated structure, we plot the integrated profile of $J_z$ across the inhomogeneous $x$ direction for the CIC-40ppc run in Fig.~\ref{fig:m180_cic_0b_cb256-multiple-Jz-profile_x-zs0}. Bifurcation starts at around $t\sim15\,\Omega_{ci}^{-1}$. As it was shown in Ref.~\onlinecite{Lee2012f} and confirmed by our simulations, it is mainly driven by a reduction of the electron current dentiy $J_{e,z}$. This in turn, it is due to a reduction of
the electron bulk velocity $V_{e,z}$, while the electron density distribution $n_{0e}$ does not vary too much ($J_{e,z}=-en_{0e}V_{e,z}$). The ion current density $J_{i,z}$ is practically unchanged.
Neglecting the heat flux, this is consistent with a relation between the pressure anisotropy (proportional to the temperature anisotropy) and the derivative of the profile of the electron bulk velocity $V_{e,z}$ (see Ref.~\onlinecite{Schindler2008})
\begin{equation}
	\frac{P_{xx,e}-P_{zz,e}}{P_{xx,e}} =\frac{1}{B_{y}}\frac{dV_{e,z}}{dx} \propto \frac{T_{e,x}-T_{e,z}}{T_{e,x}}.
\end{equation}
This relation also agrees with the behaviour seen in the TSC-40ppc run: since there is no significant electron heating, no significant reduction in $V_{e,z}$ is observed at all and, therefore, no bifurcation. And, reconnection, the last stage of the tearing island merging, occurs for TSC-40ppc but not for CIC-40ppc.

Let us now discuss the specific effects of the temperature anisotropy on the instabilities of the CS. The electron temperature anisotropy can be quantified  as $A_e=T_{e,\parallel}/T_{e,\perp}$, with $T_{e,\parallel}=T_{e,y}$ and $T_{e,\perp}=(T_{e,x}+T_{e,z})/2$ the temperatures in the directions parallel and perpendicular to the asymptotic Harris magnetic field (in $y$ direction). Note that the temperature that enters in the Eq.~\eqref{harris_equilibrium} (defining the Harris CS equilibrium) is the perpendicular one, $T_{e,\perp}$.

\begin{figure}[H]
	\centering
	\includegraphics[width=0.8\linewidth]{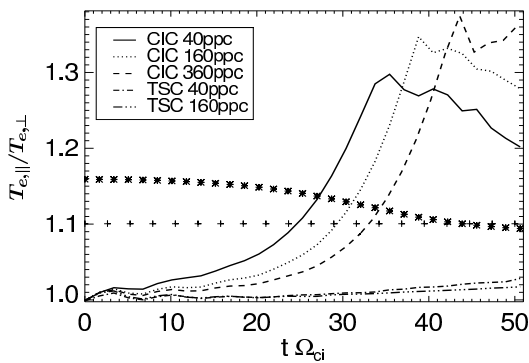}
	\caption{Time history of electron temperatures anisotropies $A_e= T_{e,\parallel}/T_{e,\perp}$ for the five (initially isotropic) runs with different shape functions and number of macroparticles per cell. The theoretical thresholds for tearing mode stabilization are depicted by $+$ Eq.\eqref{anisotropy} and $*$ Eq.\eqref{anisotropy2}. The latter one is calculated with the instantaneous value of $T_{i,\perp}/T_{e,\perp}$ for the run CIC-40ppc.}\label{fig:m180_natural anisotropies}
\end{figure}

Fig.~\ref{fig:m180_natural anisotropies} depicts the evolution of the electron temperature anisotropy  $A_e$ for the five ``initially isotropic'' runs. Note that the anisotropic heating (enhanced by using a CIC shape function) develops in the ``right`` physical direction: the parallel temperature $T_{\parallel}$ increases above the perpendicular $T_{\perp}$ during the evolution of a Harris CS. This
stabilizes the tearing mode\cite{Chen1984}. The reason for the preferential parallel heating can be easily understood: since the tearing mode develops an electric field parallel to the magnetic field, the electrons can be heated along that direction~\cite{Karimabadi2004,Vainshtein1982}. The physical anisotropic heating for the TSC-40ppc run, however, is much smaller than the numerical heating for CIC-40ppc.
All the runs which developed sufficiently strong temperature anisotropy led to bifurcation and stabilization of tearing mode.
The reason is the threshold for the tearing mode stabilization in dependence  on the electron temperature anisotropy\cite{Chen1984}. This condition can be written as\cite{Quest2010a}
\begin{equation}\label{anisotropy2}
	A_e=\frac{T_{e,\parallel}}{T_{e,\perp}}>\left(1 - 0.6\sqrt{\pi}\frac{\rho_e}{L}\left(1+\frac{T_{i,\perp}}{T_{e,\perp}}\right)\right)^{-1}.
\end{equation}
For $T_{e,\perp}\gg T_{i,\perp}$, it is possible to recover approximately the known stability criterion originally derived by Laval~\cite{Laval1967} and Forslund\cite{Forslund1968}
\begin{equation}\label{anisotropy}
	A_e=\frac{T_{e,\parallel}}{T_{e,\perp}}\gtrsim\left(1 - \frac{\sqrt{2}\rho_e}{L}\right)^{-1}.
\end{equation}
For our equilibrium parameters with $T_{i,\perp}=T_{e,\perp}$, the right hand side of Eq.~\eqref{anisotropy2} is $1.159$.  But, due to the strong electron heating, the factor $\left(1 + T_{i,\perp} / T_{e,\perp}\right)$ deviates considerably from its equilibrium value (equal to 2) in the course of the CIC shape function simulations. For that reason, we plot Eq.~\eqref{anisotropy2}   in Fig.~\ref{fig:m180_natural anisotropies} on dependence of the instantaneous value of $\left(1 + T_{i,\perp} / T_{e,\perp}\right)$  for the run with strongest heating CIC-40ppc (this is the minimum value for the threshold), instead of its equilibrium value. As it can be seen in Fig.~\ref{fig:m180_natural anisotropies}, there is a good match with the stabilization threshold for all cases  that developed bifurcation and suppressed the tearing mode.

The theoretical reason behind this stabilization (see Refs.~\onlinecite{Karimabadi2004,Karimabadi2005})
is the apparition of another instability driven by electron thermal anisotropies: the Weibel unstable mode\cite{Weibel1959,Treumann2013b}. According to the classical theory for a homogoneous plasma\cite{Krall1973}, the Weibel mode is a very low frequency electromagnetic wave instability (real frequency $\omega\sim 0$). The propagation direction and maximum growth rates of the Weibel  instability are perpendicular to the warmer temperature (associated to $T_{e,\perp}$ in our case), while it is damped and does not propagate in the colder direction (associated to $T_{e,\parallel}$. In our 2D case, the direction of Weibel damped mode turns out to be the same as the wave vector $k_y$ of the tearing mode  ($y$ or parallel $\parallel$). Both Weibel and tearing modes are driven by Landau resonance. Hence, they are coupled and as a result, the tearing mode instability becomes damped as well.

Note that our choice of parameters is especially sensitive to temperature anisotropies, since they are enhanced most for small gyroradii and sufficiently realistic (large) mass ratios~\cite{Chen1984}. It is also important to remark that temperature anisotropy effects are relevant mostly for electrons at ion time scales, comparable with $\Omega_{ci}^{-1}$. Moreover, the Weibel instability driven by ion temperature anisotropy is much weaker than that driven by electron anisotropy.
Therefore, in order to determine the threshold of tearing mode stabilization, it is not necessary to take into account ion temperature anisotropies unless they are very large and the electron temperature anisotropy is negligible small\cite{Baumjohann2010} . Since this is not the case in any of our simulations, the ion temperature anisotropy can be safely neglected for all the cases of simulations presented in this paper.
\subsection{Numerical CS bifurcation and entropy}
\begin{figure}[H]
	\centering
	\includegraphics[width=\linewidth]{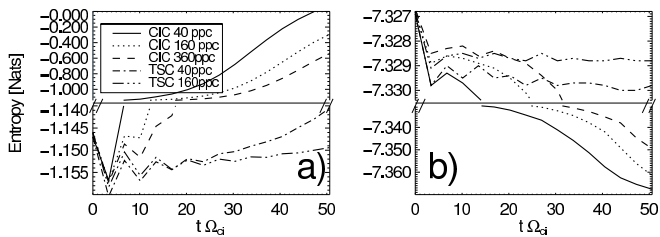}
	\caption{Time histories of electron (a) and ion (b) entropies $S(t)$ calculated according to Eq.~\eqref{entropy_histogram} for the five isotropic runs. This entropy can quantify the numerical collisions. The electron entropy curves follow the same trend as the curves for the total energy and electron temperatures shown in Fig.~\ref{fig:0b_temp_comparison}. And for the same reasons explained in that  Fig.~\ref{fig:0b_temp_comparison}, we break the $y$ axis in two halves with different scales.}\label{fig:entropies}
\end{figure}

In the course of rising  anisotropic heating  and CS bifurcation, a previous study (Ref.~\onlinecite{Lee2012f}) analyzed the entropy growth. By using the XOOPIC code~\cite{Verboncoeur1995}, that provides only the CIC shape function, it was found a growing entropy since the very beginning and its associated bifurcation later on. This was associated with the physical process of electron chaotic scattering which takes place if the heated electrons cross the center of the CS\cite{Buchner1989,Zelenyi2003}. As it was shown at the beginning of this section, our run CIC-40ppc with the ACRONYM code reproduces this bifurcation.
Regarding the entropy, in Ref.~\onlinecite{Lee2012f} the relative entropy with respect to the initial velocity distribution function, sometimes called Kullback-Leibler divergence\cite{Kullback1951}, was used. Let us use instead the conventional definition of information entropy \cite{Shannon1948},
\begin{equation}
	S(t)= -\int f(\vec{v},t)\; {\rm ln} f(\vec{v},t)\;d^3v,
\end{equation}
which differs from the thermodynamic (or Gibbs) entropy only by the Boltzmann's constant $k_B$.
This quantity is approximated by means of a histogram estimator\cite{Moddemeijer1989}
\begin{equation}\label{entropy_histogram}
	S\approx - \sum_{i,j,k}^{n_x,n_y,n_z} \rho_{i,j,k} \; {\rm ln} \rho_{i,j,k}  + 3\,{\rm ln}(\Delta v),
\end{equation}
where $\rho_{i,j,k}=f_{i,j,k}\Delta v^3$ are the bin-related probabilities of the macroparticles to occupy a cell (labelled $i,j,k$) in the velocity space $v_x$,$v_y$,$V_{e,z}$ with a bin size $\Delta v^3$. In \eqref{entropy_histogram}, $n_x$,$n_y$,$n_z$ are the number of cells in each direction of the velocity space and $f_{i,j,k}$ is the discrete approximation of the continuous distribution function $f(\vec{v},t)$ in each cell labelled $i$-$j$-$k$.
The entropy \eqref{entropy_histogram} is defined up to a constant offset depending on the relative units of $\Delta v$ (in our case, in units of the speed of light $c$). Since we are using natural logarithms, the units of $S$ are nats (1.44 bits). We checked that $S$ is more or less invariant through a wide range of choices of $\Delta v$. Thus, the time evolution of the entropy for the five isotropic runs is shown in Fig.~\ref{fig:entropies}.
Note that the ion entropy varies very little for different shape functions and number of macroparticles, since the ion time scales are much larger than those of the electrons. Hence, the ion contribution to the total entropy is negligible. On the other hand, the electron entropy for the CIC-40ppc run shows a growing behaviour, reproducing the result of Ref.~\onlinecite{Lee2012f}.
But choosing a higher order scheme for the shape function (see curve for the TSC-40ppc run) keeps the entropy more constant in time. To a lesser extent, the same effect is achieved by using more macroparticles per cell. This is in agreement with our previous results concerning the suppression of quasi-collisional heating by using higher order shape functions (see Fig.~\ref{fig:0b_temp_comparison}).

A collisionless Vlasov plasma by definition describes reversible processes and thus keeps the entropy constant for a closed system (see Ref.~\onlinecite{Klimontovich1997} for a discussion about non-equilibrium processes and the validity of the Vlasov approximation in a real plasma). But an increase in the entropy is expected in PIC codes, because the discretization of space due to the introduction of a grid and finite time stepping (the coarse-graining of phase-space\cite{Buneman1964}) can create entropy.
This is true even for a system in thermodynamic equilibrium which should have maximum entropy for some given constraints (see section 12.6 of Ref.~\onlinecite{Birdsall1991}).  The rate of entropy increase can be calculated from the Boltzmann equation with an effective collision operator given by Eq.~\eqref{collision}. The latter is proportional to the increase of kinetic energy according to
\begin{eqnarray}
	  && \frac{d}{dt}\frac{1}{2}\overline{(\vec{v}-\overline{\vec{v}})^2} = -v_{th,e}^2\frac{d S}{dt}        \\
	  && =  -\frac{1}{2v_{th,e}^2}\int d\vec{v}d\vec{v}'(v_i- v_i')Q_{ij} (v_j- v_j')\nonumber f(\vec{v})f(\vec{v}').
\end{eqnarray}
This is valid for an instantaneous Maxwellian distribution function. Here, $Q_{ij}$ is a tensor proportional to the right hand side of the numerical collision operator, therefore having the same dependence as Eq.~\eqref{collision}. And to the order and approximations used in the kinetic equation, the rate of change of kinetic energy  is equal to the rate of change of total energy. This explains the correlation between the curves for the total energy, Fig.~\ref{fig:m180_anisotropies}(a), and the electron entropy, Fig.~\ref{fig:entropies}(a) (in particular the growing curves for the CIC runs). This correlation has already been noticed in early uniform and electrostatic simulation studies (see, e.g., Ref.~\onlinecite{Montgomery1970}).

The results showed in this section can be interpreted in the following way. It is known that the entropy increases due to the presence of dissipation. This process can be characterized by the diffusion coefficient $D_{ij}$ that appears in the effective collision operator of the Boltzmann equation \eqref{collision_operator}. It was mentioned that this coefficient depends on the shape function among other parameters specific to PIC codes. Therefore, the increase of entropy in momentum-conserving PIC codes is an indicator of the strength of numerical collisions. Finally, those numerical collisions or scattering generate the non-physical heating observed in our simulations (the second numerical heating mechanism mentioned in  Sec.~\ref{sec:intro}). Note that previous investigations\cite{Okuda1970} have shown that the difference between the ideal Vlasov and Boltzmann equation, given by the numerical collision operator, is especially important for 2D3V plasma models in comparison with the 3D cases.

The numerical collisions can even provide an explanation for the location in which bifurcation takes place (at the center of the CS). In fact, in a previous work\cite{Matsuda1975} it was shown that numerical collisions in electromagnetic PIC code simulations are reduced in regions with stronger magnetic field. Since the magnetic field strength is minimum, initially even zero at the center of a Harris-type CS, we expect that the numerical quasi-collisional effects are dominant there. In this way, the enhanced numerical collisionality produces a stronger numerical heating around the center $x=0$ (not shown here), which in turns generates a bifurcated CS. This same previous work\cite{Matsuda1975} noticed that the numerical collisions are highly anisotropic in a 2D configuration with an externally applied magnetic field.
The electron-ion collisions, measured through the temperature relaxation time, drag and diffusion coefficients, depend on the relative direction of the magnetic field relative to the electric field fluctuations. They stated that is due to the neglect of the spatial variations in the $z$ direction:  a 2D3V PIC code constrains the motion of the macroparticles to the $x$-$y$ plane, but solves for all three components of the velocities $v_x$-$v_y$-$v_z$.  However, the analysis in Ref.~\onlinecite{Matsuda1975} was carried out for a plasma embedded in a strong magnetic field. Therefore, it may not apply to our case, especially near the center of the CS. For this reason we tested that behaviour by means of two small simulations with CIC shape function to enhance the numerical collisions. The first one is a 2D run with the same parameters as the run CIC-40ppc, a smaller simulation box [$128\Delta x\times128\Delta x$] and without the initialization of a Harris sheet (only a constant magnetic field in the $y$ direction with magnitude $B_{\infty y}$). And the second one is a full 3D simulation with the same previous parameters but in a box [$128\,\Delta x\times128\,\Delta x \times 128\,\Delta x$]. The 3D run developed a much smaller numerical heating and electron temperature anisotropy than the 2D one. This indicates that 2D simulations overestimate the effects of numerical collisions and thus, the origin of the anisotropy is mainly due to the neglect of 3D effects.
\subsection{Numerical heating and background plasma effects}\label{sec:background}
So far,  we did not include a background population in the previously described runs. This might have important implications in the numerical scattering, since at the edge of the CS there are large electromagnetic fluctuations due to the small number of macroparticles in that region. In order to clarify if the neglect of a background plasma is affecting our conclusions, we investigated its influence in our simulations.

\begin{figure}[H]
\centering
\includegraphics[width=\linewidth]{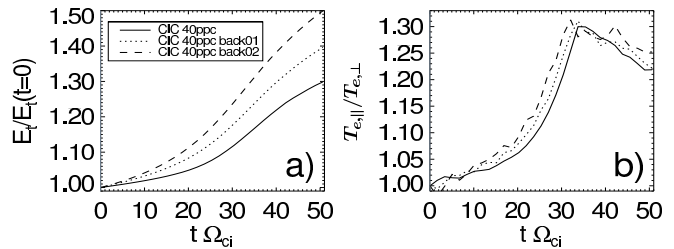}
		\caption{Time history of a) total energy (normalized to its initial value) and b) electron temperature anisotropy $A_e=T_{e,\parallel}/T_{e,\perp}$ for the cases of a CS without background (CIC-40ppc), background population at $10\%$ of the peak density (CIC-40ppc-back01) and at $20\%$ of the peak density (CIC-40ppc-back02). In all those runs the CIC scheme has been used with 40ppc.}\label{fig:background_effects}
\end{figure}

\begin{figure}[H]
\centering
\includegraphics[width=\linewidth]{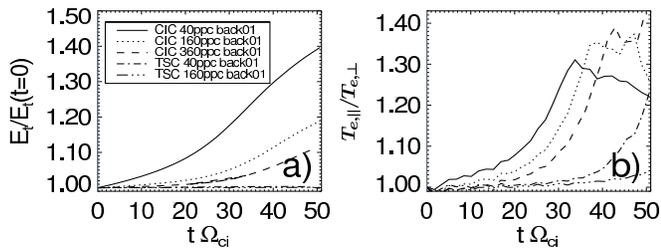}
		\caption{Time history of a) total energy and b) electron temperature anisotropy $A_e=T_{e,\parallel}/T_{e,\perp}$ for our standard runs described in Sec.~\ref{sec:heating_shapefunction}, but with the addition of  a background population of $10\%$ of the peak density. Compare with the Figs.~\ref{fig:0b_temp_comparison}(a) and \ref{fig:m180_natural anisotropies}.}\label{fig:background_effects2}
\end{figure}

For our standard runs with 40ppc and CIC scheme the results are depicted in Fig.~ \ref{fig:background_effects}. These runs are for the cases with no background (CIC-40ppc), $10\%$ (4 ppc, CIC-40ppc-back01) and $20\%$ (8 ppc, CIC-40ppc-back02) of the peak density of the Harris plasma population. All of them show a similar evolution for the temperature anisotropy. Correspondingly, we found, by looking the time history of $J_z$ profiles, that the bifurcation of the CS evolves in a  practically indistinguishable way compared to the case without background (see Fig.~\ref{fig:m180_cic_0b_cb256-multiple-Jz-profile_x-zs0}). Note, however, that the total energy  is less conserved for higher plasma background densities. This could be interpreted as an indication that the influence of the scattering is enhanced mainly due to the electron temperature anisotropy, and does not depend too much on other involved quantities that may be affecting the evolution of the total energy.
As an additional test, we plotted the evolution of the total energy and electron temperature anisotropies for simulations with the same parameters used in our standard runs shown in Fig.~\ref{fig:0b_temp_comparison}, but with the addition of a background population of $10\%$ of the peak density. The results are shown in Fig.~\ref{fig:background_effects2}. By comparing with the corresponding cases without background plasma (Figs.~\ref{fig:0b_temp_comparison}(a) and \ref{fig:m180_natural anisotropies}), one can see an almost identical evolution for the curves of the CIC scheme. Only for the TSC scheme some differences are obtained in the evolution of the electron temperature anisotropy. This is because in those cases, the explosive phase of magnetic reconnection (more accurately modeled with the second order interpolation scheme TSC) takes place a little bit earlier in the background cases than without it. This is in agreement with previous Vlasov (Ref.~\onlinecite{Schmitz2008}) and PIC (Ref.~\onlinecite{Karimabadi2005}) simulation results of magnetic reconnection. Those authors reported that the onset of magnetic reconnection and the saturation of tearing instability vary little depending on a background, while the reconnection rates are substantially reduced with an increasing background plasma density. We also observed that the explosive phase of reconnection is shorter without background than with it. This is due to the larger Alfv\'en velocity in the low density regions away from the center of the CS, where it decreases exponentially.
Therefore, our conclusion is that the addition of a background population does not seem affect too much the bifurcation/stabilization of the tearing mode in the course of the evolution of a run with a CIC scheme. This imply that nor the developed increasing temperature anisotropy nor the numerical scattering are affected too much by this population. Hence, the large fluctuations at the edge of a current sheet (due to the limited number of macroparticles in the simulations without background plasma) do not affect too much the physical processes that we modeled in this paper. This is because bifurcation takes place near the center of the current sheet, away from the regions with large fluctuations due to the too small number of macroparticles per cell. For other physical processes, the inclusion of a background may become critical for a correct description (e.g.: any estimation that requires the calculation of momenta of distribution function at the edge of the CS).
\section{Stabilization of tearing mode for initially imposed temperature anisotropies: ''initially anisotropic runs``}\label{sec:anisotropic_runs}
\subsection{Initial temperature anisotropy relaxation}
Let us investigate the cause for the relation between anisotropic heating and CS bifurcation. For this sake, we performed simulations with an initial temperature anisotropy $A_e$, obtained by choosing $T_{e,y} = T_{e,\parallel}$ below and above its equilibrium value $T_{e,\perp}$. All the other parameters are kept identical to those of the TSC-40ppc run, in order to minimize the numerical heating effects, in particular 	$T_{e,\perp}=T_{e,x}=T_{e,z}$. Note  that due to our choice of $T_i=T_e$, we also have an ion temperature anisotropy $A_i=T_{i,\parallel}/T_{i,\perp}$. These additional five runs, with temperature anisotropies $A_e=A_i=0.64$, $1.21$, $1.44$, $1.69$ and $1.96$ (corresponding to the choice of electron thermal speeds of $v_{th,ey}/v_{th,e\perp}=0.8$, $1.1$, $1.2$, $1.3$ and $1.4$, respectively), are denoted as TSC-40ppc-A0.64, TSC-40ppc-A1.21, TSC-40ppc-A1.44, TSC-40ppc-A1.69 and TSC-40ppc-A1.96, respectively. We preferred to sample more values with $A_e>1$ than $A_e<1$, because those correspond to the temperature anisotropies obtained using CIC shape functions (cf. Fig.~\ref{fig:m180_natural anisotropies}).

\begin{figure}[H]
	\centering
	\includegraphics[width=\linewidth]{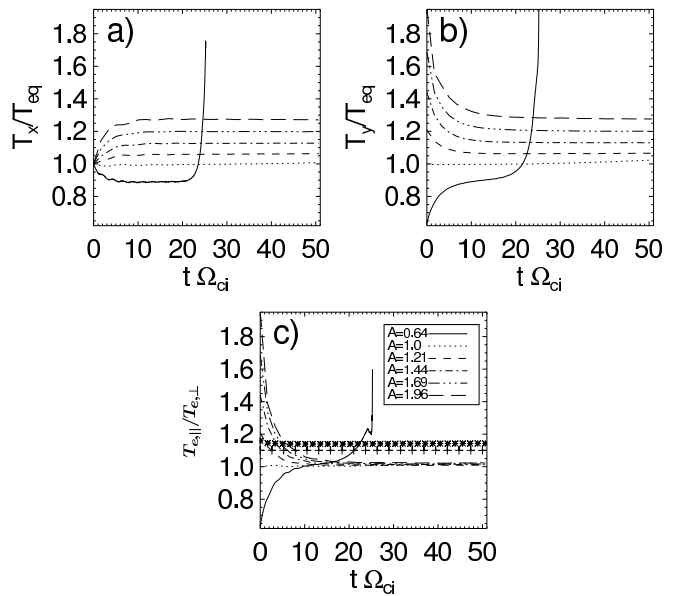}
	\caption{Time history of quantities characterizing the five runs with initially imposed anisotropy  $A_e$ (''anisotropic runs''), compared to the ``isotropic run'' TSC-40ppc. a) average electron temperatures $T_{e,x}$,  b) average electron temperatures $T_{e,y}=T_{e,\parallel}$ and c) average electron temperature anisotropy $T_{e,\parallel}/T_{e,\perp}$ for different initially imposed temperature anisotropies $A_e\in[0.6,2.0]$. The theoretical thresholds for stabilization of tearing mode are shown with lines $+$ Eq.\eqref{anisotropy} and $*$ Eq.\eqref{anisotropy2}. The latter one is calculated with the instantaneous value of $T_{i,\perp}/T_{e,\perp}$ for the run TSC-40ppc-A1.96.}\label{fig:m180_anisotropies}
\end{figure}

Some results of the ''anisotropic runs`` are illustrated in Fig.~\ref{fig:m180_anisotropies}. The parallel temperature $T_{e,y}$, shown in Fig.~\ref{fig:m180_anisotropies}(b), tends to decrease towards its equilibrium isotropic value for all the runs with $A_e>1$. The final value is proportional to the initially imposed $A_e$. On the other hand, we can see in Fig.~\ref{fig:m180_anisotropies}(a) that $T_{e,x}$ (same for $T_{e,z}$, not shown here) increases departing from its initially imposed equilibrium value,  approaching to an asymptotic value also proportional to the initial value of $A_e$. In summary, in all those cases the system relaxes towards  a more isotropic state during a short transient time, as can bee seen in Fig.~\ref{fig:m180_anisotropies}(c).
At a first sight, the results regarding the fast relaxation of the initially imposed anisotropy shown in Fig.~\ref{fig:m180_anisotropies} may seem to be a numerical artifact. This is because, as we already mentioned, the only temperature that enters into the Harris sheet equilibrium is the perpendicular one $T_{e,\perp}$. Therefore, an initial state with a different parallel temperature $T_{e,\parallel}$ is still a Harris equilibrium, and the initially imposed temperature anisotropy should be conserved in an ideal collisionless Vlasov system without additional instabilities (although this equilibrium turns out to be unstable to an anisotropy driven instability under the conditions to be described). Hence, in order to prove the validity of the runs shown in Fig.~\ref{fig:m180_anisotropies}, we carried out two simulation runs for the case with the highest considered anisotropy (TSC-40ppc-A1.96), while changing the number of macroparticles per cell:  TSC-360ppc-A1.96 (360 macroparticles per cell) and TSC-1000ppc-A1.96 (1000 macroparticles per cell). Those runs have, respectively, three and five times less numerical noise compared to TSC-360ppc-A1.96, since (as we already mentioned) the noise level is reduced according to $\sqrt{N}$, with $N$ the number of macroparticles per cell.
\begin{figure}[H]
	\centering
	\includegraphics[width=0.70\linewidth]{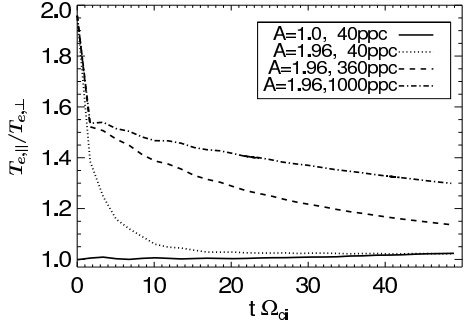}

	\caption{Time history of the average electron temperature anisotropy $A_e=T_{e,\parallel}/T_{e,\perp}$ for cases with an initially imposed temperature anisotropy $A_e=1.96$, TSC shape function and different number of macroparticles per cell (40, 360 and 1000).  For comparison, the result for an isotropic initial distribution ($A_e=1.0$) is also plotted with a solid line.}\label{fig:anisotropy_ppc}
\end{figure}
One can see in Fig.~\ref{fig:anisotropy_ppc} that in case of more macroparticles per cell, the electron temperature anisotropy  $A_e$ does not fall as quickly as for the case TSC-40ppc-A1.96, especially for later times.  There is a trend (convergence) towards the reduction of the relaxation of temperature anisotropy for an increasing number of macroparticles per cell (see, e.g., the curve for the run TSC-1000ppc-A1.96 in Fig.~\ref{fig:anisotropy_ppc}). This behaviour can be attributed to the numerical collisions, since they enhance the isotropization of the distribution function (even by using the TSC scheme) and are weaker for a larger number of macroparticles per cell. This process is developed via pitch angle scattering, redistributing the particle velocities in the momentum space: a momentum transfer takes place from the parallel velocity direction to the perpendicular one, with a negligible change of the particle energy (total speed).
\subsection{Physical origin of the initial temperature anisotropy relaxation: Weibel instability}\label{sec:weibel}
\begin{figure}[H]
\centering
	\includegraphics[width=\linewidth]{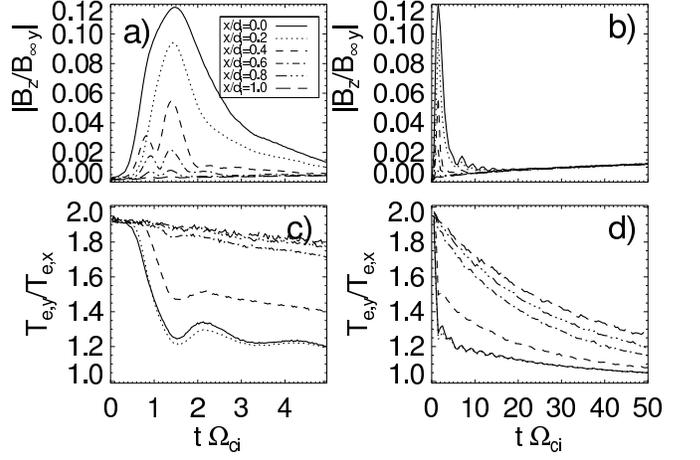}
\caption{Time history of the out of plane magnetic field and temperature anisotropy averaged along $y$, at different distances $x$ from the center of the CS, for the run TSC-360ppc-A1.96. a) and b) Magnetic field  $|B_z|$. c) and d) Electron temperature anisotropy $T_{e,y}/T_{e,x}$. The left plots a) and c) are for the early stages of the system while the right plots b) and d) are for the long term evolution.}\label{fig:evolution_bz_ani}
\end{figure}

In spite of the previous numerical argument, the initial drop in $A_e=T_{e,\parallel}/T_{e,\perp}$ shown in the curves of Fig.~\ref{fig:m180_anisotropies}(c) or Fig.~\ref{fig:anisotropy_ppc}  (especially for high values of $A_e$) seems to be independent on the number of macroparticles per cell, indication of a physical origin. In order to understand this process, it is convenient to look more closely the evolution of the electron temperature anisotropy in this initial stage, focusing in the case TSC-360ppc-A1.96. This is shown in Fig.~\ref{fig:evolution_bz_ani}
We can distinguish three stages in the evolution of the system: (1) the initial between $t\Omega_{ci}=0-0.5$ in which the anisotropy is constant, (2) between $t\Omega_{ci}=0.5-1.5$ in which the anisotropy is quickly relaxed and (3) $t\Omega_{ci}>1.5$, where the system relaxes its anisotropy slowly until very late times.

In the initial stage (1), due to the high value of the initial imposed anisotropy $T_{e, y}>T_{e, x},T_{e, z}$, the unmagnetized center of the CS becomes unstable to the Weibel instability. The timescales in which Weibel operates are much shorter than these of the tearing mode, and therefore the aforementioned coupling between these instabilities can be neglected in this early stage (as well as any consequence of this coupling, such as  the stabilization of tearing mode). We already mentioned that this instability propagates in the colder direction associated to $T_{e,\perp}=T_{e,x}$  (since no wave vectors in $z$ direction are allowed in our 2D configuration). Its growth rate can be approximated as
\begin{equation}\label{weibel_gamma}
	\gamma = \sqrt{\frac{2}{\pi}}k_xv_{th,e}\frac{T_{e,x}}{T_{e,y}}\left(\frac{T_{e,y}}{T_{e,x}}-1-\left(\frac{k_xc}{\omega_{pe}}\right)^2\right).
\end{equation}
This expression is valid in the low frequency regime ($\omega/(k_xv_{th,e})\ll 1$). We have also neglected the ion contribution, since is negligible by a factor of $m_i/m_e$. Therefore, this is a instability mainly driven by electrons. From the previous expression it is possible to determinate the threshold:
\begin{equation}\label{weibel_threshold}
\left(\frac{T_{e,y}}{T_{e,x}}-1\right)  > \left(\frac{k_xc}{\omega_{pe}}\right)^2.
\end{equation}
Although this electromagnetic instability was originally discovered and described for an unmagnetized plasma, it can also be triggered under the influence of a background magnetic field (in this context, it is sometimes called ``filamentation instability''\cite{Bret2005, Bret2009}), as long as its magnitude is not too high\cite{Tautz2007}. Indeed, this kinetic instability only depends on the bulk properties of the plasma (anisotropy), and operates under a wide range of conditions due to the fact that does not depend on wave-particles resonances effects\cite{Tautz2006}. The Weibel instability generates stationary magnetic fields (which might be important, e.g., for the reconnection process\cite{Treumann2010}) perpendicular to the direction of the ``warmer'' temperature, (in our 2D setup, the $\hat{z}$ direction) by the well known mechanism described in Ref.~\onlinecite{Fried1959, Medvedev1999}. The addition of a strong background magnetic field (in our case, in the $\hat{y}$ direction) tends to inhibit this instability because the motion of the particles perpendicular to it is suppressed, in such a way that it is not possible the momentum transfer between the parallel and perpendicular directions. Under the assumption $\Omega_{ce} \ll \omega_{pe}$ (valid in the weak magnetic fields close the center of the CS), it is known that the Weibel growth rates given by Eq.~\eqref{weibel_gamma} are not modified to a good approximation. The background magnetic field only has an influence in the apparition of a small real frequency $\omega\sim \Omega_{ce}$ (see Refs.~\onlinecite{Hededal2005, Stockem2006, Tautz2008} for further details).

Therefore, the generation of the magnetic field $B_z$ (see Fig.~\ref{fig:evolution_bz_ani}(a)) close to the center of the CS, as well as the corresponding decrease in the electron temperature anisotropy (see Fig.~\ref{fig:evolution_bz_ani}(c)) are the main signatures of the development of the Weibel mode. For a quantitative comparison, we need a value for the wavevector $k_x$ in the threshold condition Eq.~\eqref{weibel_threshold}. The $k_{x,max}$ corresponding to the maximum growth rate given by Eq.~\eqref{weibel_threshold} cannot be applied directly, since it may predict longer wavelengths than the ones allowed by the unmagnetized condition of the Weibel instability $\Omega_{ce} \ll \omega_{pe}$. Indeed, according with Eq.~\eqref{weibel_gamma}, $k_{x,max}$ can be calculated as\cite{Krall1973}:
\begin{equation}\label{weibel_k_max}
	k_{x,max} = \frac{\omega_{pe}}{c}\sqrt{\frac{1}{3}\left(\frac{T_{e,y}}{T_{e,x}}-1\right)},
\end{equation}
with the corresponding maximum growth rate
\begin{equation}\label{weibel_gamma_max}
	\gamma_{max}=\sqrt{\frac{8}{27\pi}}\omega_{pe}\frac{v_{th,e}}{c}\frac{T_{e,x}}{T_{e,y}}\left(\frac{T_{e,y}}{T_{e,x}}-1\right)^{3/2}.
\end{equation}
We can note that the associated $\lambda_{x,max}=2\pi/k_{x,max}$ gets larger for smaller anisotropies. For our parameters, the initial value $A_e=1.96$ corresponds to $\lambda_{x,max}=0.8d_i$, but  the value at saturation $A_e\sim1.2$ corresponds to $\lambda_{x,max}=1.8d_i$. This length would reach strongly magnetized regions, suppressing the Weibel mode. In Ref.~\onlinecite{Lu2011}, it was proposed that Weibel instability will saturate when the wavelength along $x$ is of the order of magnitude of the halfwidth $L$ of the CS (measured from the center), since it is the typical lengthscale in which the magnetic field varies. Assuming that reasonable wavelength $\lambda\sim 2L\sim 1d_i$ (to be justified \emph{a posteriori}), we can estimate the right hand side of Eq.~\eqref{weibel_threshold} as $\left(k_xc/\omega_{pe}\right)^2\sim0.21$. This value matches well with the observed saturation anisotropy after $t\Omega_{ci}\gtrsim 1.5$. Note that this value for the saturation of Weibel mode is higher than the one required for the stabilization of tearing mode (compare to Eqs.~\eqref{anisotropy2} and \eqref{anisotropy}). This means that a system unstable to Weibel mode will already be stable to tearing mode, inhibiting reconnection. And viceversa, a PIC simulation that develops high enough numerically generated temperature anisotropy along all the center of the CS will stabilize tearing mode before of reaching a state unstable to Weibel mode (this is what happens in the previously discussed CIC runs).

\begin{figure}[H]\centering
\includegraphics[width=0.97\linewidth]{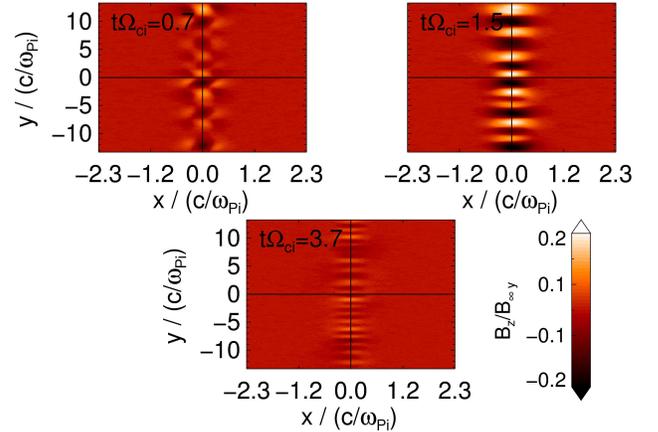}
\caption{Contours of the out of plane magnetic field $B_z$ for the run TSC-360ppc-A1.96. The snapshots correspond to three characteristic times during the evolution of the Weibel instability}\label{fig:structure_bz}
\end{figure}

It is possible to validate the previous estimation for the saturation wavelength of the Weibel mode by looking at the spatial structure of $B_z$, displayed in Fig.~\ref{fig:structure_bz}.
The ``chessboard`` (filamented) structure around the center of the CS has already been observed in previous studies of magnetic reconnection, in both electron-positron plasmas (Refs.~\onlinecite{Swisdak2008,Zenitani2008a,Liu2009}) as well as in electron-proton plasmas (Refs.~\onlinecite{Lu2011,Schoeffler2013}). In those works, it has been identified as a characteristic signature of the Weibel mode generated by the temperature anisotropy inside of the magnetic tearing islands. In this context, the anisotropy is caused by the particles bouncing inside of those structures, i.e.:, the well known Fermi mechanism\cite{Schoeffler2013}. The Weibel instability has also been proposed to be generated inside of the electron inertial length around the X point, as a result of the electron inflow \cite{Baumjohann2010,Treumann2010}.

At the beginning, when the anisotropy is high, the size of the $B_z$ structures across $x$ is relatively small,  as can be seen in Fig.~\ref{fig:structure_bz}(a). They reach larger sizes later, when the anisotropy is reduced, in agreement with Eq.~\eqref{weibel_k_max} (see Fig.~\ref{fig:structure_bz}(b)). Their saturation size at $t\Omega_{ci}\sim1.5$ is of the order of $\lambda=1d_i$, as expected from our previous estimation. They cannot grow further because of the constrain in the magnetic field strength.

Regarding the size of the $B_z$ structures along $y$, the usual linear theory for a homogeneous plasma cannot predict their length scale (there should be no wavevector along the warmer $y$ direction). But a mechanism proposed in Ref.~\onlinecite{Liu2009} can explain and predict the finite $k_y$ observed in Fig.~\ref{fig:structure_bz}. A wavevector along the direction of higher temperature can be produced with a wavelength of spatial scales that matches the double of the electron gyroradius based on the $B_z$ generated around the neutral point (see also Ref.~\onlinecite{Schoeffler2013}). At the saturation time of the Weibel instability, the average component of the magnetic field is about $B_z\sim 0.1B_{\infty y}$, implying a gyroradius $\rho_{e,z}/d_i\sim 0.35$. This is about 1/4 of the observed size of the the Weibel structures at that stage ($1.3d_i$), consistent with the predictions of Ref.~\onlinecite{Liu2009}. This mechanism also explains why for later times (see Fig.~\ref{fig:structure_bz} for $t\Omega_{ci}=3.7$) the filamented Weibel structures in $y$ direction are smaller than for $t\Omega_{ci}=1.5$, since the generated magnetic field $B_z$ becomes weaker after the relaxation of the temperature anisotropy. This implies a smaller electron gyroradius on this component and thus a smaller size of the wave vector along the $y$ direction.

\subsection{Consequence of the initial temperature anisotropy relaxation: Mirror instability}\label{sec:mirror}
After $t\Omega_{ci}\gtrsim 1.5$, the magnetic fields generated by the Weibel instability start to slowly be dissipated, as can be seen in Figs.~\ref{fig:structure_bz}(c), \ref{fig:evolution_bz_ani}(a) and \ref{fig:evolution_bz_ani}(b). This in an indication of the turning off of this instability, since the Weibel mode generates turbulent magnetic fields\cite{Tautz2008} that scatters the particles in a such a way that those tend to reduce the anisotropy in the distribution function (the source of their free energy). For later times, the anisotropy at the center of the CS $x=0$ tends to be kept at the saturation values, although with a slow decrease (see Fig.~\ref{fig:evolution_bz_ani}(d)). From this same figure it is possible to see that the low density regions further away from the center, not affected in a significant way by the Weibel instability, will experience a a faster isotropization in comparison with the center (due to the enhanced collisions). Note that this effect cannot be seen in the average of the electron temperature anisotropy (see Fig.~\ref{fig:anisotropy_ppc}).

Although the Weibel instability seems to be active until $t\Omega_{ci}\sim 1.5$, there is an important difference in the behaviour of the system after $t\Omega_{ci}\sim 0.5$. From that particular time onwards, a secondary instability driven by temperature anisotropy starts to develop around the neutral point in the CS: the so called mirror instability\cite{Hasegawa1969,Gary1993}. Different from the unmagnetized Weibel instability, this electromagnetic instability requires a magnetic field in order to exist. This is precisely the magnetic field provided by the Weibel instability. The mirror instability is one of the instabilities that can be excited in a plasma when $T_{j,\perp}>T_{j,\parallel}$, for both ions (classical hydromagnetic ion mirror mode\cite{Gary1993}) and electrons\cite{Gary2006}. Its threshold condition can be written as \cite{Hall1979,Pokhotelov2000,Hellinger2007}:
\begin{equation}\label{mirror_threshold}
	\sum_j \beta_{j,\perp}\left(\frac{T_{j,\perp}}{T_{j,\parallel}} -1 \right)>1 + \frac{\left(\sum_j \frac{T_{j,\perp}}{T_{j,\parallel}}\right)^2}{2\sum_j \beta_{j,\parallel}^{-1}},
\end{equation}
or, more explicitly,
\begin{eqnarray}\label{mirror_threshold2}
	D=&&\left(\frac{T_{i,\perp}}{T_{i,\parallel}} -1\right) +\frac{T_{e,\perp}}{T_{i,\perp}}  \left(\frac{T_{e,\perp}}{T_{e,\parallel}} -1\right) - \frac{1}{\beta_{i,\perp}}\\
	&&-\frac{T_{e,\parallel}T_{i,\parallel}}{2T_{i,\perp}(T_{e,\parallel} + T_{i,\parallel})}\left(\frac{T_{i,\perp}}{T_{i,\parallel}} - \frac{T_{e,\perp}}{T_{e,\parallel}}\right)^2>  0,\nonumber
\end{eqnarray}
where the plasma betas are  $\beta_{j,\parallel}=\frac{n_jk_BT_{j,\parallel}}{B^2/(2\mu_0)}$ and $\beta_{j,\perp}=\frac{n_jk_BT_{j,\perp}}{B^2/(2\mu_0)}$. Here, the parallel ($\parallel$) and perpendicular ($\perp$) directions are with respect to the local magnetic field. The well known mirror threshold condition derived from fluid equations\cite{Hasegawa1975,Treumann2001a} assumes cold electrons, resulting in a vanishing last term in \eqref{mirror_threshold} or \eqref{mirror_threshold2} (this term is a correction due to the field aligned electric field arising from the differential motion of ions and electrons\cite{Pokhotelov2000}). However, we checked that this last term does not play an important role in our this system, since initially is imposed to be zero due to our initialization (in addition, it saturates at high enough electron temperatures).  It is important to notice that the threshold condition \eqref{mirror_threshold} involves the contribution of both ions and electrons. In our originally anisotropic runs, both species are anisotropic and contributes in more or less the same proportion to the threshold.  The mirror modes propagates at an oblique angle with respect to the magnetic field (maximum growth rate nearly in the perpendicular direction), and have predominant longitudinal fluctuations\cite{Treumann2001a}. One key feature of the mirror instability is its preferential development in high $\beta$ plasmas with relatively low anisotropies\cite{Genot2001}, since the threshold is lower (see Eq.\eqref{mirror_threshold2}). Therefore, a weakly Weibel magnetized center of a CS (high $\beta$) with an imposed temperature anisotropy has the ideal conditions for the triggering of this instability\cite{Zenitani2008a}.

It is important to note that the (electron) whistler instability is other instability driven by the same condition $T_{e,\perp}>T_{e,\parallel}$. But we ruled out the existence of that instability in this system, since its maximum growth rate is in the parallel direction to the magnetic field (in this case, mostly in the $\hat{z}$ direction), which is not allowed in our 2D simulations.

\begin{figure}[H]\centering
\includegraphics[width=0.97\linewidth]{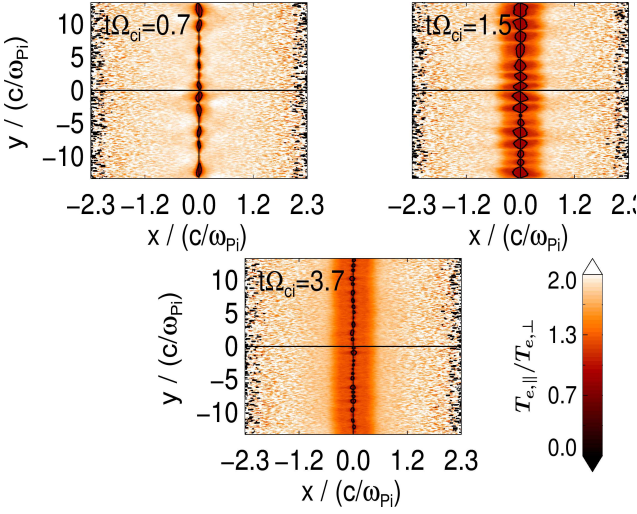}
\caption{Contours of the electron temperature anisotropy $T_{e,\parallel}/T_{e,\perp}=P_{\parallel,e}/P_{\perp,e}=\beta_{\parallel,e}/\beta_{e,\perp}$ for the run TSC-360ppc-A1.96. The snapshots correspond to the same three times shown in Fig.~\ref{fig:structure_bz}. Note that regions with $T_{e,\perp}>T_{e,\parallel}$ are shown inside of the black contour lines.}\label{fig:structure_temps}
\end{figure}

\begin{figure}[H]\centering
\includegraphics[width=0.7\linewidth]{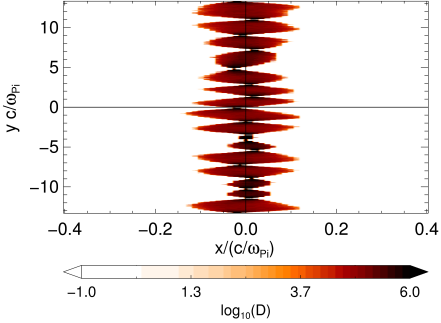}
\caption{Contours of the mirror threshold value from Eq.~\eqref{mirror_threshold2}: $\log_{10}(D)$. Only positive values are shown (corresponding to instability). This plot is for the run TSC-360ppc-A1.96 at the time $t\Omega_{ci}=1.5$, corresponding to the peak of activity of both Weibel and mirror instabilities. Compare with the corresponding plots for $B_z$ in Fig.~\ref{fig:structure_bz} and temperature anisotropy $T_{e,\parallel}/T_{e,\perp}$ in Fig.~\ref{fig:structure_mirror}. Note that only a small region close to the current sheet is shown, in order to visualize the mirror structures that have a maximum typical length scale of $c/\omega_{pe}$ across the $x$ direction.}\label{fig:structure_mirror}
\end{figure}

\begin{figure}[H]\centering
\includegraphics[width=0.92\linewidth]{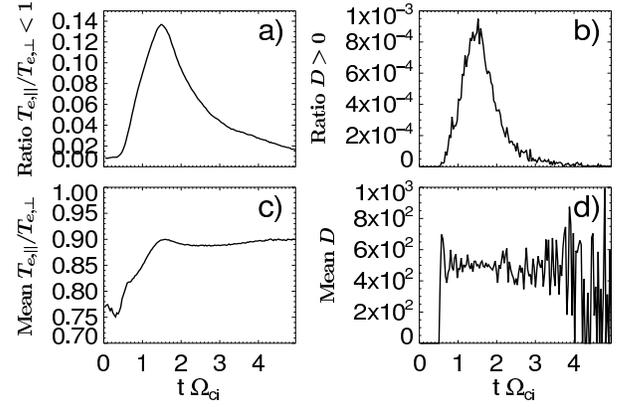}
\caption{Time evolution of quantities related with the mirror instability for the run TSC-360ppc-A1.96. a) Number of grid points in $|x|<L$ that satisfy $T_{e,\parallel}/T_{e,\perp}<1$ divided over the total in this region. Note that $T_{e,\parallel}/T_{e,\perp}$ is the electron temperature anisotropy aligned with the local magnetic field, i.e.: the $\parallel$ direction is in the $\vec{B}/|\vec{B}|$ direction.  The choice of the $|x|<L$ is basically because encompass all the area that becomes unstable to mirror instability b) Mean value of $T_{e,\parallel}/T_{e,\perp}$ for the grid points that satisfy the condition $T_{e,\parallel}/T_{e,\perp}<1$. c) Number of grid points in $|x|<L$ that satisfy the mirror threshold condition Eq.~\eqref{mirror_threshold2}, divided over the total in this region. d) Mean value of the mirror threshold condition (left hand side of Eq.~\eqref{mirror_threshold2}) calculated for the points that satisfy that condition.}\label{fig:evolution_mirror_counts}
\end{figure}

Since that around the center of the CS sheet the magnetic topology is complex (not only a $B_z$ is generated, but also a less stronger $B_x$ component), in order to calculate Eq.~\eqref{mirror_threshold} properly, it is necessary to rotate the tensor pressure $P_{ij}$ to keep it always locally aligned with the magnetic field. This means that the $\parallel$ direction is imposed to be in the $\vec{B}/|\vec{B}|$ direction. In this way, we can see that small scale islands (with a maximum typical length scale of $\sim c/\omega_{pe}$ across the $x$ direction) are formed around the center of the current sheet with $T_{e,\perp}>T_{e,\parallel}$ (see Fig.~\ref{fig:structure_temps}), where the mirror condition Eq.~\eqref{mirror_threshold} is fulfilled (see Fig.~\ref{fig:structure_mirror}).
The time evolution of this process can be seen in Fig.~\ref{fig:evolution_mirror_counts}, showing the ratio of grid points unstable to the mirror instability in a region $|x|<L$ and their mean value, as well as the condition $T_{e,\parallel}/T_{e,\perp}<1$ that can trigger this instability (number of points that satisfy this condition divided over the total in $|x|<L$). We can see that the regions unstable to mirror instability start to appear only after $t\Omega_{ci}=0.5$. They reach  their peak (both in size and strength) around $t\Omega_{ci}=1.5$, correlated with the maximum of the Weibel generated magnetic field $B_z$ and the minimum in the temperature anisotropy $T_{e,y}/T_{e,x}$. Between those times, the source of the free energy of the mirror mode, the field aligned temperature anisotropy $T_{e,\perp}/T_{e,\parallel}>1$, is quickly exhausted (see Fig.~\ref{fig:evolution_bz_ani}). The mirror instability produces strong turbulence in those unstable regions, deforming and kinking the in-plane magnetic field lines (since it acts mainly in the perpendicular direction to the local magnetic field). The mirror structures only survive for short times. Note that the Weibel instability by itself cannot explain the sudden appearance of the small scale structures seen only between $0.5\lesssim t\Omega_{ci}\lesssim 3$.
Other signatures that can support our conclusion that those structures are due to the mirror instability (in addition to the threshold condition) can be provided by a comparison with the linear theory. However, due to the coupling with the Weibel instability acting at the same time, it is not easy a direct comparison, especially in our case of non homogeneous plasma with the mirror instability only acting in a very narrow region. Nevertheless, the linear theory can provide us with some insight. In this context is meaningful to write the usual expression for the maximum growth rate of the mirror mode  \cite{Hasegawa1975,Treumann2001a}, assuming cold electrons and an ion anisotropy:
\begin{eqnarray}\label{mirror_gamma}
\gamma=&&\sqrt{\frac{2}{\pi}}\frac{T_{i,\parallel}}{T_{i,\perp}\beta_{i,\perp}}\Bigg[\beta_{i,\perp}D\Bigg. \\
&&+\left.\frac{k_{\parallel}^2}{k_{\perp}^2}\sum_j \left(\frac{\beta_{j,\parallel}-\beta_{j,\perp}}{2} -1 \right) \right]k_{\parallel}v_{th,i,\parallel},
\end{eqnarray}
where $D$ is the left hand side of the mirror threshold condition given by Eq.~\eqref{mirror_threshold2}. Because the condition of almost perpendicular propagation, the second term is neglected for the mirror mode (it gives the growth rate of the complementary instability: the so called firehose instability driven by $T_{j,\parallel}>T_{j,\perp}$ with parallel propagation). A more precise growth rate taking into account similar electron and ion contributions to the temperature and anisotropy of the system (more suitable for our case) was given in Ref.~\onlinecite{Pokhotelov2000}:
\begin{align}\label{mirror_full}
\gamma_{max}=\frac{k_{\perp}v_{th,i,\parallel}T_{i,\parallel}}{\sqrt{\pi}T_{i,\perp}\beta_{i,\perp}}\frac{4A^{3/2}\left(1 + \frac{T_{e,\parallel}}{T_{i,\parallel}}\right)^2\left[1 + (\beta_{i,\perp}-\beta_{i,\parallel})/2\right]}{\left(1 +
\frac{T_{e,\parallel}}{T_{i,\parallel}}\right)^2 + \left(1 + \frac{T_{e,\perp}}{T_{i,\perp}}\right)^2 },
\end{align}
with
\begin{equation}\label{mirror_afactor}
A =\frac{D}{3\beta_{i,\perp}^{-1}\left(1+\frac{\beta_{i,\perp}-\beta_{i,\parallel}}{2}\right)}.
\end{equation}
Although this expression increases with $k_{\perp}$, it is only valid in the small Larmor radius approximation $k_{\perp}\rho_i\ll1$. Following Ref.~\onlinecite{Pokhotelov2000}, the instability growth rate will decrease in the short wavelength limit due to finite Larmor radius effects, and therefore would reach a maximum around the ion gyroradius $k_{\perp}^{-1}\sim \rho_i$. In Fig.~\ref{fig:structure_mirror}, the longitudinal size of the spatial structures (tilted with respect to the $y$ direction), where the mirror condition is satisfied, agrees well with this prediction. Indeed, a typical structure has a size in $l_x\times l_y=l_{\parallel}\times l_{\perp}=0.2d_i\times2.0d_i$, while $\lambda_{\perp,max}=2\pi/\rho_i \sim 3d_i$.

On the other hand, the mirror mode predicts a small propagation vector along the parallel direction. Although most of the magnetic field has $z$ component, there is a small contribution from the $x$ component. Since our 2D geometry does not allow wavevectors in the out-of-plane direction $z$, the parallel propagation direction has to be taken necessarily by the $x$ direction (although the contribution for the parallel temperature comes mostly of $T_{e,z}$). Therefore, the mirror mode will propagate mostly in the perpendicular $y$ direction in such a way that $k_{\perp}=k_{y}\gg k_{\parallel}=k_{x}$. The angle $\theta$ of propagation (with respect to the magnetic field) can be estimated by the following expression:
\begin{equation}\label{angle}
\tan(\theta)=\frac{k_{\perp}}{k_{\parallel}}=\frac{1}{\sqrt{A}}.
\end{equation}
For typical parameters in our simulation inside of the mirror structures ($\beta_{i,\parallel}\sim \beta_{e,\parallel}\sim80$, and the field aligned anisotropies $A_e\sim A_i\sim0.99$), this angle is around $\theta\sim 85^\circ$, which agrees very well with the tilting angle of a typical mirror structure ($\theta={\rm atan(2.0d_i/0.2d_i)}\sim 85^\circ$). Note that although the range of variation of the parameters inside the magnetic mirror structures can be very large, especially regarding the plasma betas, the angle $\theta$ have a weak dependence on those parameters (mirror modes usually have nearly perpendicular propagation), and does not deviate too much from this value (different from the growth rate, that can be very sensitive to those variations).
\subsection{Growth rates of the initial temperature anisotropy relaxation due to Weibel/mirror instabilities}
\begin{figure}[H]\centering
\includegraphics[width=0.95\linewidth]{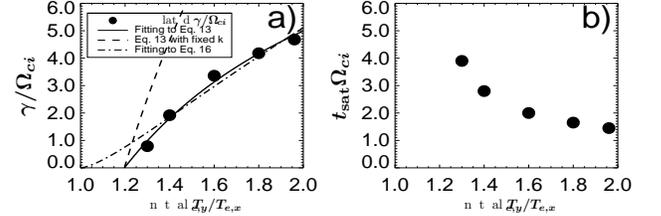}
\caption{a) Comparison among simulated growth rates of the Weibel instability during the linear stage on dependence of the initially imposed electron temperature anisotropy $T_{e,y}/T_{e,x}$, a fitting according to Eq.~\eqref{weibel_gamma} (for fixed $k$), a fitting according to Eq.~\eqref{weibel_gamma_max} (for $k_{x,max}$), and the theoretical growth rates calculated from the initial parameters in Eq.~\eqref{weibel_gamma} (for fixed $k$). See text for details. b) Saturation time $t_{sat}$ of the Weibel instability on dependence of the initially imposed electron temperature anisotropy $T_{e,y}/T_{e,x}$.}\label{fig:fitting}
\end{figure}

In order to compare growth rates of both Weibel and mirror instabilities, we used the results of runs with different levels of initially imposed anisotropy and 360ppc. For lower values of the initial anisotropy, the peak value of $B_z$ is lower and it is reached for later times. Therefore,  by performing a linear fit in the plots of the mean value of $|B_z(x=0)|$ between the times in which this magnetic field grows exponentially, we can measure the Weibel-mirror mode growth rates. The results are shown in Fig.~\ref{fig:fitting}. As we already mentioned, when the initial anisotropy is lower than $A_e\lesssim 1.2$, the Weibel mode is stabilized and it cannot generate the magnetic field necessary for the formation of those structures.

In Fig.~\ref{fig:fitting}, the fitting by continuous line was calculated by using  $\gamma(A_e)/\Omega_{ci}=(c_1/A_e)\times(A_e-1-c_2)$, equal to the functional dependence of the analytical growth rates given by Eq.~\eqref{weibel_gamma} for the Weibel instability, assuming a fixed $k$. The resulting fitting coefficients are $c_1=12.3$ and $c_2=0.20$. This implies a $k_x$ corresponding to a wavelength $\lambda_x=1.0 d_i$, in good agreement with the maximum size of the structures seen in the out of plane magnetic field $B_z$. The other fitting in dashed-point line was calculated by using  $\gamma(A_e)/\Omega_{ci}=(c_3/A_e)\times(A_e-1)^{3/2}$, equal to the functional dependence of the analytical growth rates given by Eq.~\eqref{weibel_gamma_max} for the Weibel instability  assuming a $k_{x,max}$ dependent on the temperature anisotropy, as predicted by the linear theory. The resulting fitting coefficient is $c_3=10.22$.  It can be clearly seen that this curve does not reproduce well enough the behaviour of the simulated growth rates.  On the other hand, the predicted growth rates obtained by applying directly Eq.~\eqref{weibel_gamma} for our parameters and the previous fixed $k_x$ (corresponding to the maximum size of the observed structures in $B_z$), do not agree with the observed ones (see dashed line in Fig.~\ref{fig:fitting}). The last ones are about three times lower than those given by the linear theory Eq.~\eqref{weibel_gamma}. There are many reasons for this disagreement, since we are clearly applying this theory beyond its limits (e.g.: assumptions of homogeneity, weak magnetic field, combination with mirror instability).  Therefore, we conclude that the linear theory for a homogeneous plasma can only predict the right functional dependence on $A_e$ for the growth of the Weibel instability (combined with mirror), even in our complex geometry of a Harris current sheet (see the fitting with continuous line in Fig.~\ref{fig:fitting}). However, the latter is valid only under the assumption of choosing beforehand a  fixed $k$ that depends on the typical length scale of variation of both magnetic field and density (double of the halfwidth $2L\sim 2\rho_i\sim 1.0\, d_i$).

On the other hand, the mirror growth rates \eqref{mirror_gamma} or \eqref{mirror_full} are very sensitive to the value of $\beta_i$, whose value can vary over some orders of magnitude inside of the mirror structures and so is the range of variation of the predicted $\gamma$. Therefore, the value of $\gamma$ of the mirror instability inside of the structures does not provide information useful to be compared, at least in an easy way, with any of the large scale electromagnetic fields or momenta of the distribution function observed in our simulations.

So, all the previous arguments seems to indicate that, in addition to the Weibel instability, a mirror instability is also acting close to the center of the current sheet driven by the electron temperature anisotropy. Previous studies have indicated the close relationship between the mechanism that triggers those instabilities~\cite{Hasegawa1969, Pokhotelov2010}. Especifically, in Ref.~\onlinecite{Pokhotelov2010} it was noted that the linear responses of the distribution function for the mirror-type and Weibel-type perturbations, in addition to the linear growth rates, have the same form, replacing the quantities where the ion/electron gyroradius is involved (in the mirror case) with the electron inertial length (in the Weibel case).

It is also is important to mention that this interplay between Weibel and mirror instabilities has already been pointed out explicitly in Ref.~\onlinecite{Treumann2014}. These authors concluded that in a originally homogeneous unmagnetized but anisotropic plasma, the magnetic fields generated by the Weibel instability can produce mirror modes structures. Those will be ordered into a chain of bubbles or holes, in addition to the filamented structures due to Weibel. The conclusion is that  a homogeneous but anisotropic plasma will naturally evolve into a turbulent state due to the interaction of those instabilities (see Ref.~\onlinecite{Treumann2014} and references therein).

We cannot rule out completely the presence of other temperature anisotropy driven instabilities in our system. Initially, the regions away from the center of the CS satisfy the condition to be firehose unstable. This is (mostly) a parallel propagation instability driven by the opposite condition to the mirror instability $T_{e,\parallel}>T_{e,\perp}$\cite{Krall1973,Gary1993,Camporeale2008}. But we have not found signatures of its presence (e.g.: absence of resonant protons, no clear wave vector in those regions). The reasons seems to be that its predicted growth rates are about one order of magnitude lower than those of the Weibel mode. Besides of that, all the phenomena related with bifurcation happens close to the center of the CS, especially the reduction in the temperature anisotropy, where this kind of parallel propagating instabilities should not play a role. However, it is important to remark that, under the right conditions, the firehose instability can also be triggered in a current sheet, producing kinking of the magnetic field lines, scattering particles and reducing the temperature anisotropy\cite{Matteini2013}, in a similar way to the phenomena observed in our case.
\subsection{Suppression of tearing instability by bifurcation due to anisotropic heating: ''initially  anisotropic runs``.}
The main consequence of an initial electron temperature anisotropy $A_e\gtrsim1.2$ is that, after a short transition period, a bifurcated CS is formed. This value corresponds to the one calculated in Sec.~\ref{sec:weibel} for the stabilization of the Weibel instability, assuming the mentioned constraint in $k_x$ due to the inhomogeneity in the background Harris magnetic field. Note that this value is higher than the theoretical threshold for the stabilization of tearing mode (Eqs.~\eqref{anisotropy2} and \eqref{anisotropy}). Indeed, in Fig.~\ref{fig:m180_anisotropies}(c), it is shown the threshold for stabilization of the tearing mode according to Eq.~\eqref{anisotropy2}. Since the ratio $ T_{i,\perp} / T_{e,\perp}$ does not change too much away from its equilibrium values, the threshold curve is almost horizontal. Therefore, a system unstable to Weibel mode not only will be stable to the tearing mode (as we mentioned in Sec.~\ref{sec:weibel}), but also will develop bifurcated structures in $J_z$.

\begin{figure}[H]
		\centering
		\includegraphics[width=\linewidth]{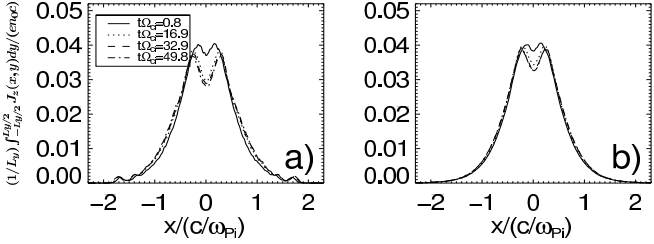} %
	\caption{Evolution of the total current density profile $J_z$ showing bifurcation for runs with the highest considered initially imposed temperature anisotropy $A_e=1.96$, TSC shape function and different number of macroparticles per cell: a) TSC-40ppc-A1.96 and b) TSC-360ppc-A1.96. The profiles are obtained by integrating the current density along the CS: $J_z(x) = (1/L_y)\int_{-Ly/2}^{Ly/2} J_z(x,y) dy$.
	}\label{fig:multiple-anisotropies-comparison}
\end{figure}

\begin{figure}[H]\centering
\includegraphics[width=0.97\linewidth]{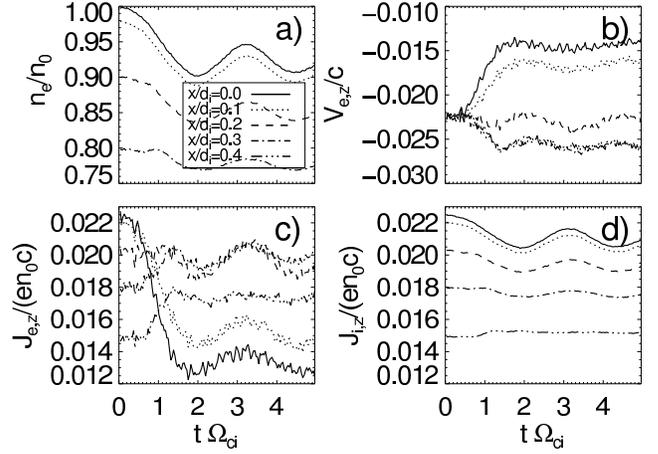}
\caption{(Short) time evolution for some quantities related with bifurcation, averaged along $y$, at different distances $x$ from the center of the CS, for the run TSC-360ppc-A1.96. a) Electron density $n_e$. b) Electron bulk speed $V_{e,z}$. c) Electron current density $J_{e,z}$. d) Ion current density $J_{i,z}$. We do not show the ion density $n_i$ since its values in the course of the evolution are always very close to $n_e$, i.e.: the quasineutrality condition is kept everywhere to a large extent.}\label{fig:evolution_bifurcation}
\end{figure}

The morphology of the bifurcation in $J_z$ is similar to the resulting from evolving temperature anisotropy (see Fig.~\ref{fig:m180_cic_0b_cb256-multiple-Jz-profile_x-zs0}).
The bifurcation is stronger (i.e.: the dip in the current density is deeper) for higher initial temperature anisotropies. See Fig.~\ref{fig:multiple-anisotropies-comparison} for a comparison of this behaviour in the evolution of the profile $J_z(x)$ for the runs with highest considered anisotropy and different number of macroparticles per cell: TSC-40ppc-A1.96 and TSC-360ppc-A1.96.
Note that the initially imposed anisotropy is still causing a bifurcation of the CS for the cases with more macroparticles per cell and, correspondingly, still stabilizes the tearing mode instability. However, the drop in $J_z$ at the center of the CS in TSC-360ppc-A1.96  is less deep compared with the case  TSC-40ppc-A1.96. This indicates a trend towards the reduction of bifurcation in the limit of negligible scattering.
In order to understand the onset of the bifurcated structures, it is necessary to recall the results concerning the relaxation of the initial anisotropy by the two instabilities previously discussed. In particular, we have already mentioned that the Weibel instability plays a critical role in the development of the tearing mode. Therefore, it is expected that the mirror instability  can also be important in the stability properties of the tearing mode. Indeed, it has already been identified that the force responsible for both anisotropic tearing and mirror instability have very similar origins\cite{Chen1984,Shi1987}. And since we saw a correlation between stabilization of tearing mode and formation of bifurcated structure in the CIC runs, it is worthwhile to analyze the relation between bifurcation and mirror instability happening as a consequence of an imposed temperature anisotropy.
In order to do that, let us investigate the time evolution of the quantities related with bifurcation at different distances from the center of the current sheet. The results are shown in Fig.~\ref{fig:evolution_bifurcation}.
We can see that the region close to the center of the CS, unstable to the mirror instability from $t\Omega_{ci}\gtrsim 0.6$, also develops a drop in both density $n_e$ and the out-of-plane electron bulk velocity $V_{e,z}$. This leads to a decrease in the out of plane current density $J_z$. The drop in density due to an increase in the magnetic pressure is a known signature of the mirror mode\cite{Southwood1993},  that can be explained in a fluid approach (the fluctuations in $n_0$ and $B_0$ should be in anti-phase). But the drop in $V_{e,z}$ is due to a different process: the pitch angle scattering that is active during that time. The bifurcation is stopped when the anisotropy is reduced, and so is the scattering. This can be proven by means of a comparison of the previous Fig.~\ref{fig:multiple-anisotropies-comparison} with  Fig.~\ref{fig:m180_cic_0b_cb256-multiple-Jz-profile_x-zs0}. The latter shows that bifurcation obtained by the CIC scheme increases as the time goes by due to the corresponding increase of the electron temperature anisotropy. But in that CIC case, the scattering is not only of pitch-angle type, but also due to the overall heating with the associated increment in the total particle energy.

It is of central importance to note that even after the stabilization of the mirror mode, both $n_e$ and $V_{e,z}$ remain with the minimum values reached during the brief time of strong mirror interactions, even for very long times. The mirror structures are not static, but they oscillates around the center of the CS. The consequence of this process is the formation, after the mirror structures disappears, of a narrow strip region where the current density $J_z$ is reduced. The neighborhood of this region that never became mirror unstable preserves the original value of $J_z$. As as final result, we observe a bifurcated current structure: a drop of $J_z$ at the center of the CS (see Fig.~\ref{fig:multiple-anisotropies-comparison}). We can even explain to what extent the bifurcation will modify the structure of the current density, since its maximum extension has to be proportional to the component of the mirror wavevector in the parallel direction $k_{\parallel}$. Assuming a fixed $k_{\perp}$ (see explanation in Sec.~\ref{sec:mirror}), this quantity has to be proportional to $\sqrt{A}$ by Eq.~\eqref{angle}. Then, using Eq.~\eqref{mirror_afactor} and \eqref{mirror_threshold2}, it turns out that this the extension across $x$ of the mirror structures is directly proportional to the instantaneous value of the anisotropy $A_e^{-1}-1$, and so is the extension of bifurcation (as can be expected). Inside of the mirror structures, the average field aligned temperature anisotropy never deviates too much from the marginal stability condition, and this explain the bifurcation happening only very close to the center of the CS.
\subsection{Influence of the number of macroparticles per cell on the initial  Weibel/mirror instabilities}
\begin{figure}[H]
\centering
	\includegraphics[width=\linewidth]{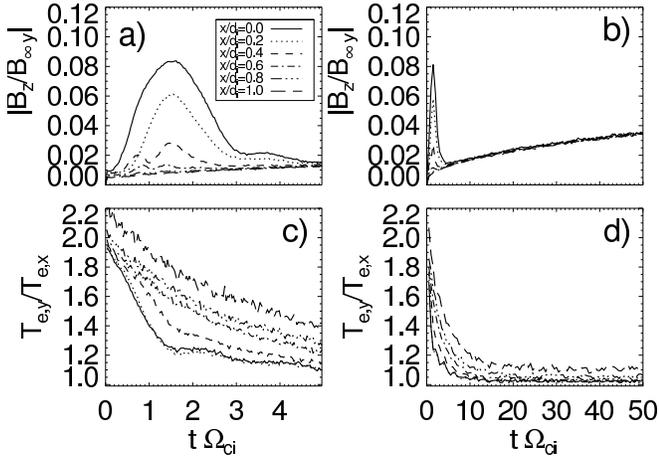}
\caption{Time history of the out of plane magnetic field and temperature anisotropy averaged along $y$, at different distances $x$ from the center of the CS, for the run TSC-40ppc-A1.96. a) and b) Magnetic field  $|B_z|$. c) and d) Electron temperature anisotropy $T_{e,y}/T_{e,x}$. The left plots a) and c) are for the early stages of the system while the right plots b) and d) are for the long term evolution. To be compared with Fig.~\ref{fig:evolution_bz_ani}.}\label{fig:evolution_bz_ani_40ppc}
\end{figure}

In Fig.~\ref{fig:anisotropy_ppc}, we can see a difference in the development of the global anisotropy by using different number of macroparticles per cell, especially in the long term evolution. The initial stage seems to be similar, but it has a critical difference. This can be shown, for the case with lowest number of macroparticles and highest anisotropy TSC-40ppc-A1.96, through plots similar to those of Fig.~\ref{fig:evolution_bz_ani}. The results are in Fig.~\ref{fig:evolution_bz_ani_40ppc}.
We can see that the initial electron temperature anisotropy $T_{e,y}/T_{e,x}$ is relaxed from the very beginning: there is no the ''plateau'' between $0<t\Omega_{ci}<0.5$ as seen in the Fig.~\ref{fig:evolution_bz_ani}(c) (for the quantities at the center of the CS $x=0$).  As a result, the free energy stored in the anisotropy can be converted from the very beginning in magnetic energy due to the Weibel instability, at a faster rate than with 360ppc (compare Fig.~\ref{fig:evolution_bz_ani}(a) with Fig.~\ref{fig:evolution_bz_ani_40ppc}(a)). For the points close to the center of the CS, this can be attributed to the enhanced noise level for 40ppc. This noise can provide a higher initial seed magnetic field from which the Weibel generated magnetic fields can grow faster. In this context, it is interesting to note that the asymptotic value reached by $B_z$ for very long times is independent of the distance to the center of the CS, being only dependent on the number of macroparticles per cell.

\begin{figure}[H]\centering
\includegraphics[width=0.97\linewidth]{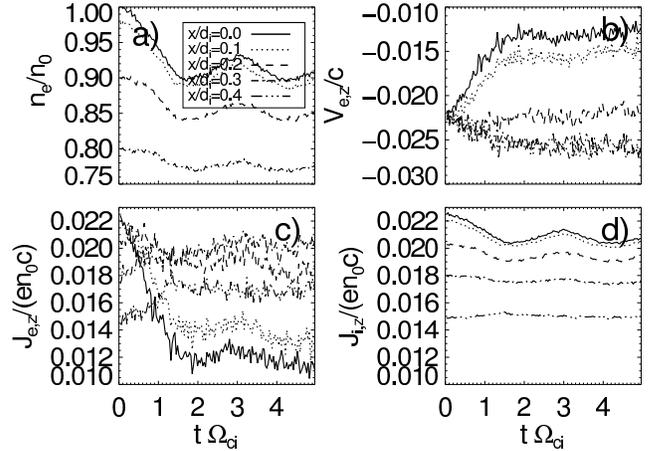}
\caption{(Short) time evolution for some quantities related with bifurcation, averaged along $y$, at different distances $x$ from the center of the CS, for the run TSC-40ppc-A1.96. a) Electron density $n_e$. b) Electron bulk speed $V_{e,z}$. c) Electron current density $J_{e,z}$. d) Ion current density $J_{i,z}$. We do not show the ion density $n_i$ since its values in the course of the evolution are always very close to $n_e$, i.e.: the quasineutrality condition is kept everywhere to a large extent. To be compared with Fig.~\ref{fig:evolution_bifurcation}}\label{fig:evolution_bifurcation_40ppc}
\end{figure}

The consequence of the enhanced growth of magnetic field $B_z$ in the early stages (for this case TSC-40ppc-A1.96) is a faster generation of the conditions for the triggering of the mirror instability at the center of the CS. We checked that the temperature anisotropy aligned with this generated magnetic field reaches $T_{e,\parallel}/T_{e,\perp}<1$ earlier than for the case with TSC-360ppc-A1.96. However, the Weibel and mirror instabilities saturate more or less at the same time, once this initial anisotropy is exhausted. This implies that the center of the CS experiences more time unstable to the mirror instability, and that can explain why bifurcation is stronger (deeper drop in $J_z$) in the cases with less number of macroparticles per cell. This can be shown by comparing the time evolution of bifurcation related quantities TSC-360ppc-A1.96 in Fig.~\ref{fig:evolution_bifurcation} with the corresponding plots for TSC-40ppc-A1.96 shown in Fig.~\ref{fig:evolution_bifurcation_40ppc}.
\subsection{Influence of ions in the temperature anisotropy driven instabilities}
In order to prove that stabilization of tearing mode and bifurcation are only due to electron temperature anisotropy, we also tested the effects of ion temperature anisotropy by means of two other sets of anisotropic simulation runs (results not shown here). In the first set, we isolated the effects of electron temperature anisotropy by imposing initially isotropic ions $A_i=1$ and anisotropic electrons $A_e\neq 1$ (with $A_e\in[0.64,1.96]$), while all the other parameters were identical to the isotropic run TSC-40ppc. We found practically the same behaviour as shown in Fig.~\ref{fig:m180_anisotropies} with both anisotropic species: the existence of a threshold for the stabilization of the Weibel instability and the subsequent generation of mirror modes structures and bifurcation (associated with the stabilization of tearing mode). The values are very similar to those ones found previously, but not exactly the same. This is because our choice of parameters implies an initial ratio $T_{i,\parallel}/T_{e,\parallel}\neq 1$, that favours different processes related with the Landau damping: the damping of some wave modes depends on $T_{i,\parallel}/T_{e,\parallel}$, as well as the thresholds and growth rates of Weibel/mirror instabilities.

For the second set of anisotropic runs, we isolated the effects of ion temperature anisotropy by imposing $A_i\neq1$ and isotropic electrons $A_e=1$ (with $A_i\in[0.64,1.96]$), while all the other parameters are identical to the isotropic run TSC-40ppc. As expected, we found that the ion temperature anisotropy does not change too much in the course of a typical simulation, since their time scales are much longer than electron time scales. The CS only develops a slow growth of tearing mode for the same timescales as the isotropic run TSC-40ppc. There is only a slow trend towards isotropization of temperatures, that can be entirely explained by the pitch angle scattering caused by an insufficient number of macroparticles per cell. As can be expected, the CS success to bifurcate only for values of $A_i$ much larger than those found for $A_e$. Regarding the tearing mode, this fact agrees with the theoretical prediction\cite{Quest2010a}: an ion temperature anisotropy  will stabilize the tearing mode growth only if it is very large. This is, however, unlikely to be caused by numerical heating, which mostly affects the electrons.

The absence of anisotropy driven instabilities in the case with only anisotropic ions is because the Weibel instability requires mostly an electron anisotropy to be triggered, rather than ion anisotropy.
Therefore, with isotropic electrons there is no generated magnetic field $B_z$ that can provide the necessary conditions for the triggering of the  the mirror instability: a field aligned temperature anisotropy due to magnetization and rotation of the magnetic field at the center of the CS. As a consequence, there is no bifurcation of the CS at all. Note that if we would have magnetic fields as those provided with Weibel instability, but with only anisotropic ions, a bifurcated structure should also be produced. This is because the mirror threshold also depends on the ion anisotropy, in more or less the same extent compared to the electron anisotropy (see Eq.~\eqref{mirror_threshold2}).
\subsection{Enhancement of tearing mode with the opposite anisotropy}
Now, let us discuss the run TSC-40ppc-A0.64  with an initial electron temperature anisotropy  $A_e=T_{e,\parallel}/T_{e,\perp}=0.8<1$ in the opposite direction to the previously discussed. In this case, the tearing instability is faster than for the isotropic temperature case. The explosive phase of the tearing instability takes place already at $t\sim20\,\Omega_{ci}^{-1}$. This explains the sudden growth of the temperature and anisotropy seen in Fig.~\ref{fig:m180_anisotropies} for this run: the quantities  $T_{e,\parallel}$, $T_{e,\perp}$ and $T_{e,\parallel}/T_{e,\perp}$  go off the scale beyond $t\gtrsim20\,\Omega_{ci}^{-1}$. From the faster merging of magnetic islands follows that the tearing mode is enhanced for temperature anisotropies $A_e<1$, in agreement with Ref.~\onlinecite{Chen1984}.

All the results shown in this section confirm that the electron temperature anisotropy $A_e$ can lower the growth rates of the tearing mode instability or even stabilizes it completely if it is strong enough. And it may also cause bifurcation due to the always non negligible noise and associated numerical scattering present in PIC simulations of this physical system.
\section{Influence of numerical parameters on the tearing mode instability}\label{sec:tearing}
\subsection{Influence of the shape function}
\begin{figure}[H]
	\centering
	\includegraphics[width=0.90\linewidth, bb = 0 20 504 360]{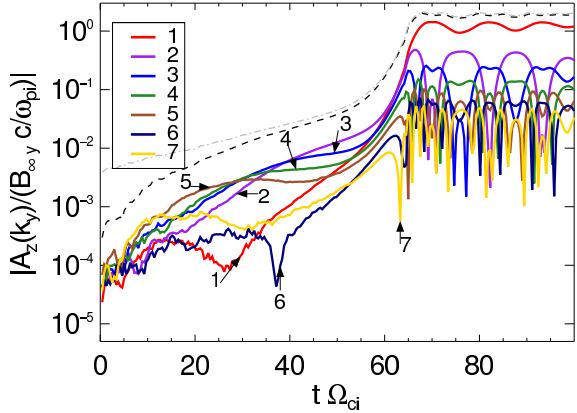}
	\caption{Time history of the first seven Fourier modes  $\int_{x=-\Delta_{Ns}}^{x=\Delta_{Ns}}|A_z(x,k_y)|\,dx$ for the run TSC-40ppc. The dashed gray line is obtained as a sum of all Fourier modes.  The dashed-dotted black line is the sum of the first seven Fourier modes shown in this plot.}\label{fig:m180_tsc_0b_cb256-fft_az_timehistory}
\end{figure}

Tearing mode magnetic islands, regions with closed magnetic field lines, are formed very early in the CS evolution. They are due to the initial numerical noise and are ordered in a long chain along the CS in $y$ direction.
Afterwards, those structures start to merge via a coalescence process due to the attraction of the current pinches, generating larger islands (up to widths of the order of the ion inertial length $d_i=c/\omega_{pi}$, marginally stable according to the linear theory). This is a characteristic signature of the tearing instability (see Fig.~\ref{fig:bifurcation}(b)). The evolution of this process can be estimated by integrating the Fourier modes of the vector potential $|A_z(x,k_y)|$ in the electron singular layer thickness $\pm \Delta_{NS}=\sqrt{2\rho_e L}\;$ \cite{Katanuma1980}. In this region, most of the electrons are unmagnetized and experience meandering instead of gyromotion~\cite{Galeev1984}. With our choice of parameters, seven linearly unstable tearing modes satisfying $k_yL\leq 1$ are possible (shorter wavelength modes are evanescent).
The time history of those modes for the TSC-40ppc run are shown in Fig.~\ref{fig:m180_tsc_0b_cb256-fft_az_timehistory}. In that figure we also overplotted the sum of all Fourier modes (gray) and the sum of the first seven ones (black). Obviously, there is a discrepancy between both curves at early times. This means that modes with shorter wavelengths than mode number seven ($k_yL>1$) contribute in a significant way to the total power only in the initial transient stage. Later on, they can be completely neglected.
\begin{figure}[H]
	\centering
	\includegraphics[width=0.90\linewidth, bb = 0 20 504 360]{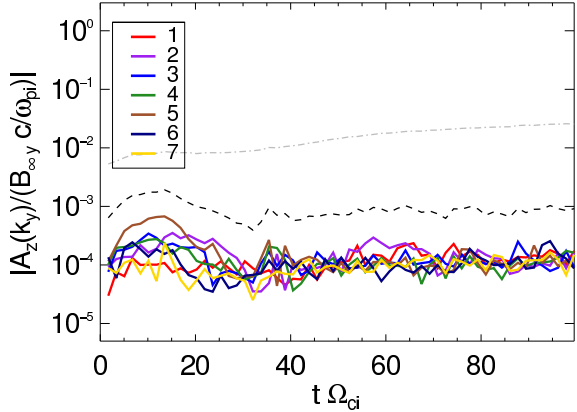} %
	\caption{Time history of the first seven Fourier modes  $\int_{x=-\Delta_{Ns}}^{x=\Delta_{Ns}}|A_z(x,k_y)|\,dx$ for the run CIC-40ppc. The dashed gray line is obtained as a sum of all Fourier modes.  The dashed-dotted black line is the sum of the first seven Fourier modes shown in this plot.}\label{fig:m180_cic-fft_az_timehistory}
\end{figure}
As expected, all tearing modes in Fig.~\ref{fig:m180_tsc_0b_cb256-fft_az_timehistory} reach the saturated stage at the same time as the reconnected flux (cf. Fig.~\ref{fig:m180_shape_fluxes}(b)). But the time history of the Fourier modes for the run CIC-40ppc, Fig.~\ref{fig:m180_cic-fft_az_timehistory}, reveals a complete suppression of the tearing modes, correlated with the respective non-growing reconnected flux (see Fig.~\ref{fig:m180_shape_fluxes}(a)).
This is due to the stabilizing effect of bifurcation~\cite{Camporeale2005a,Matsui2008a} that dominates the whole structure of the CS. It is also interesting to notice that short wavelength modes with $k_yL>1$ carry a significant amount of power for most of the evolution of the CS, in contradiction with the linear theory. From this, we conclude that it is essential to compute growth rates of the tearing instability with the TSC shape function instead of the often used CIC scheme, unless a very large number of macroparticles per cell is used.
\subsection{Comparison with linear theory}
\begin{figure}[H]
	\centering
	\includegraphics[width=0.90\linewidth, bb = 0 20 504 360]{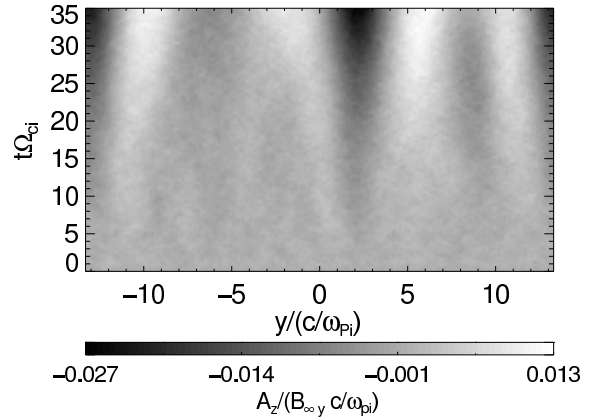}
	\caption{Time history of the profiles of vector potential $A_z(x=0,t)$ up to the limit of linear stage of tearing mode evolution ($t\Omega_{ci}\sim35$), for the run TSC-40ppc.
		}\label{fig:m180_tsc_0b_cb256-stacked_profiles_az_short}
\end{figure}

\begin{figure}[H]
	\centering
	\includegraphics[width=0.90\linewidth, bb = 0 20 504 360]{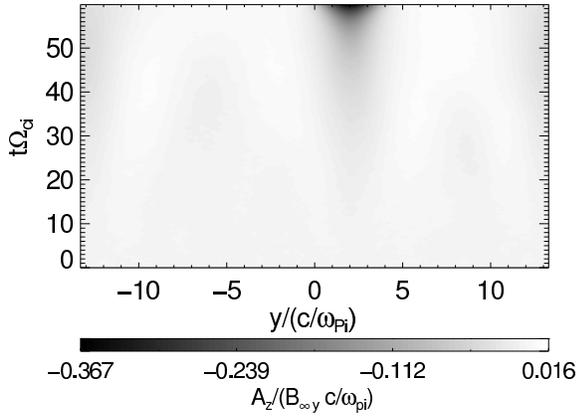}
	\caption{Time history of the profiles of vector potential $A_z(x=0)$ up to the explosive phase of reconnection ($t\Omega_{ci}\sim60$), for the run TSC-40ppc.
		}\label{fig:m180_tsc_0b_cb256-stacked_profiles_az_long}
\end{figure}

In order to compare the linear growth rates of the tearing mode with the linear predictions, it is necessary first to identify the linear stage of the instabilities in the simulation results. This is not straightforward in multimode tearing, because a complex transfer of energy between different modes takes place at different times.
We chose the following criterion: the linear regression is performed up to the time at which it starts to be noticeable a fast growth of the short-wavelength modes $k_yL> 1$ ($M\geq7$) that should be stable  according to the linear theory (we did not show those modes with  practically vanishing growth rates in Fig.~\ref{fig:m180_tsc_0b_cb256-fft_az_timehistory}). For later times, the system enters into a non-linear stage, and thus a comparison with linear theory is not possible anymore.
A comparison between linear and non-linear stages can be seen in Figs.~\ref{fig:m180_tsc_0b_cb256-stacked_profiles_az_short} and \ref{fig:m180_tsc_0b_cb256-stacked_profiles_az_long}. In those figures, the time history of the vector potential along the center of the CS, $A_z(x=0)$, is displayed up to the limit of the linear stage ($t\sim35\,\Omega_{ci}^{-1}$) and up to the start of the explosive phase of reconnection ($t\sim60\,\Omega_{ci}^{-1}$), respectively.

Fig.~\ref{fig:m180_tsc_0b_cb256-stacked_profiles_az_short} shows a clear dominance of four magnetic islands (their O points are the maxima of $A_z$ and the X points the minima) at the end of the linear stage. This is in agreement with the linear theory~\cite{Pritchett1991}: the most unstable mode should have $M=4$ or $k_yL\sim0.545$ (green line in Fig.~\ref{fig:m180_tsc_0b_cb256-fft_az_timehistory}), implying that the dominant islands should have a size $H_s$ in $y$ direction of about $H_s/(c/\omega_{pi})=2\pi/(k_y\;c/\omega_{pi})=6.66$ (see, e.g., chp. 7 in Ref.~\onlinecite{Bellan2006}, chp. 12  in Ref.~\onlinecite{Biskamp2000}, Ref.~\onlinecite{Karimabadi2005a}). However, the reconnected flux between the X and O points of each one of those magnetic islands is very low: only marginally exceeds the numerical noise level. So, magnetic reconnection it is not energetically important during this stage.
Later, in the nonlinear stage, a transfer of power to the long wavelengths modes takes place. As can be seen in  Fig.~\ref{fig:m180_tsc_0b_cb256-stacked_profiles_az_long}, at the end the whole system is dominated by only one tearing island, with a much larger reconnected flux than during the linear stage. This is just before of the explosive phase of reconnection, when the entire structure of the CS is disrupted.
\begin{figure}[H]
	\centering
	\includegraphics[width=0.8\linewidth]{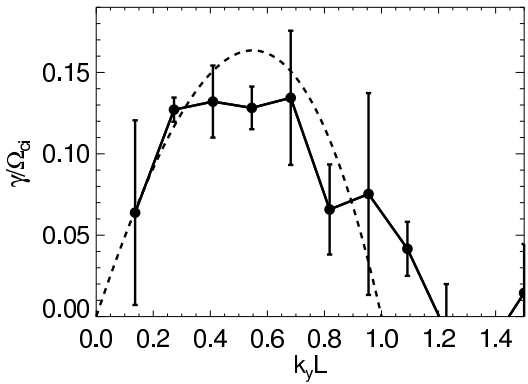}
	\caption{Dots connected by the solid line: simulated growth rates $\gamma$ of the Fourier modes  $\int_{x=-\Delta_{Ns}}^{x=\Delta_{Ns}}|A_z(x,k_y)|\,dx$ (shown in Fig.~\ref{fig:m180_tsc_0b_cb256-fft_az_timehistory}) vs wave number $k_yL$ for the run TSC-40ppc (see text for the calculation method). Dashed line: analytical estimate of the growth rate according to Eq.~\eqref{tearing_growth} for thin CS and $m_i=m_e$.
		}\label{fig:m180_tsc_0b_cb256-gamma-k}
\end{figure}
For the previously described linear stage, we compute the growth rates $\gamma$ of the tearing instability as function of  $k_yL$ by means of a linear regression of the integrated Fourier modes $\int_{x=-\Delta_{Ns}}^{x=\Delta_{Ns}}|A_z(x,k_y)|\,dx$ shown in Fig.~\ref{fig:m180_tsc_0b_cb256-fft_az_timehistory}. The results are shown in Fig.~\ref{fig:m180_tsc_0b_cb256-gamma-k}. Some modes with $k_yL>1$ are also included, in order to prove that those are linearly damped. Although the criterion for choosing the linear stage provides a first order approximation, is not good for all modes since some of them reach a non-linear stage before. For that reason, we chose several windows for performing the linear fit from very early and up to $t\sim35\,\Omega_{ci}^{-1}$. Each one provides a different growth rate, being closer to each other for a exponential growth, and we plot the average of those values in Fig.~\ref{fig:m180_tsc_0b_cb256-gamma-k}. The error bars are the standard deviation of those values. In the Fig.~\ref{fig:m180_tsc_0b_cb256-gamma-k}, the dashed lines are from an analytical formula for thin CS ($\rho_i/L\sim1$,  Ref.~\onlinecite{Pritchett1991}):
\begin{eqnarray}\label{tearing_growth}
	\frac{\gamma}{\Omega_{ci}}  =&&\frac{2\sqrt{2}}{\sqrt{\pi}}\left(\frac{\rho_i}{L}\right)^{3}\frac{k_yL(2+k_yL)(1-k_yL)}{1+4\left(\frac{\rho_i}{L}\right)^{2}} \\
	                            =&&0.2591\,k_yL\,(2+k_yL)(1-k_yL).\nonumber
\end{eqnarray}
This expression for the collisionless ion tearing mode growth rate was derived within the Vlasov formalism by matching internal and external solutions for the perturbed vector potential with respect to the electron singular layer thickness $\Delta_{NS}$. Note that this approximation to the tearing mode growth rate considers equal masses for both ion and electrons $m_i/m_e=1$. Numerical solutions of the linearized Vlasov equation~\cite{Brittnacher1995,Daughton1999,Silin2002} have shown that for realistic ion to electron mass ratios ($m_i/m_e=1836$), the growth rates are reduced by a factor of 1.5 to 2 (see section 10.4 of Ref.~\onlinecite{Schindler2007}). For the intermediate mass ratios used in our simulations, as expected, the growth rates are reduced by a smaller amount.
This explains, in part, the difference between the maximum value of the growth rate predicted by the analytical formula $ \gamma / \Omega_{ci} =0.163$ for the most unstable mode $k_yL\sim0.545$. But the main reason for the discrepancy in the overall simulated curve with the theoretical one is due to the coupling between the unstable modes for the multimode tearing, as it has been reported by previous works (see, e.g.: Ref.~\onlinecite{Karimabadi2005}).
\subsection{Influence of the number of macroparticles on tearing mode growth rates}
\begin{figure}[H]
	\centering
	\includegraphics[width=0.90\linewidth, bb = 0 20 504 360]{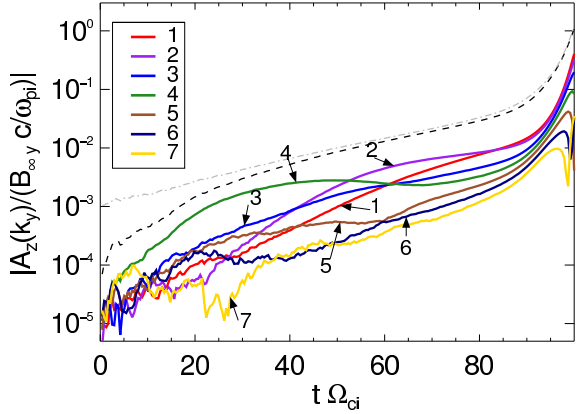}
	\caption{Time history of the first seven Fourier modes  $\int_{x=-\Delta_{Ns}}^{x=\Delta_{Ns}}|A_z(x,k_y)|\,dx$ for the run TSC-640ppc. To be compared with Fig.~\ref{fig:m180_tsc_0b_cb256-fft_az_timehistory}, for the run TSC-40ppc with sixteen times less macroparticles per cell. The dashed gray line is obtained as a sum of all Fourier modes. The dashed-dotted black line is the sum of the first seven Fourier modes shown in this plot.}\label{fig:m180_tsc_640ppc_0b_cb256_steps-fft_az_timehistory}
\end{figure}

\begin{figure}[H]
	\centering
	\includegraphics[width=0.8\linewidth]{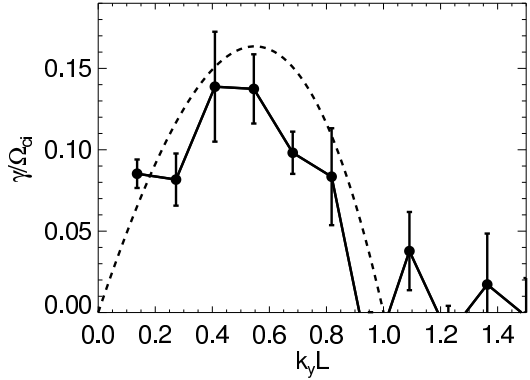}
	\caption{Dots connected by the solid line: Simulated growth rates $\gamma$ of the Fourier modes  $\int_{x=-\Delta_{Ns}}^{x=\Delta_{Ns}}|A_z(x,k_y)|\,dx$ (shown in Fig.~\ref{fig:m180_tsc_640ppc_0b_cb256_steps-fft_az_timehistory}) vs wave number $k_yL$  for the run TSC-640ppc (see text for the calculation method). Dashed line: analytical estimate of the growth rate according to Eq.~\eqref{tearing_growth} for thin CS and $m_i=m_e$. Compare with the run TSC-40ppc (sixteen times less macroparticles per cell) shown in Fig.~\ref{fig:m180_tsc_0b_cb256-gamma-k}.
		}\label{fig:m180_tsc_640ppc_0b_cb256_steps-gamma-k}
\end{figure}

In spite of all the previous comparison with the linear theory, it is important to remark that the dominant modes with higher power,  as well as the values of the growth rates, vary with the number of macroparticles used in the simulation.
There are two main reasons for this: many non-linear effects absent in the linear theory are enhanced by the numerical noise. Also, the dominance of the fastest growing Fourier modes is delayed by the same reason\cite{Matsui2008}. Thus, it is expected to have a better match with the theoretical value for a larger number of macroparticles, i.e.: lower noise level. However, in agreement with the linear theory, the most unstable mode with higher growth rates, $M=4$, is always the same. In order to
prove this claim, we run an additional simulation denoted as TSC-640ppc, with the same parameters as TSC-40ppc but with 16 times more macroparticles per cell. The latter reduces the initial noise to 1/4 of its original value. The results for the time history of the integrated Fourier modes are shown in Figs.~\ref{fig:m180_tsc_640ppc_0b_cb256_steps-fft_az_timehistory}, while the respective growth rates are in Fig.~\ref{fig:m180_tsc_640ppc_0b_cb256_steps-gamma-k} (calculated with the same previously explained method).
Comparing the evolution of the Fourier modes in Fig.~\ref{fig:m180_tsc_640ppc_0b_cb256_steps-fft_az_timehistory}  and Fig.~\ref{fig:m180_tsc_0b_cb256-fft_az_timehistory}, one sees a clear dominance (higher power) of the mode $M=4$ ($k_yL\sim0.545$) in case of more macroparticles per cell. A comparison for the growth rates in Fig.~\ref{fig:m180_tsc_640ppc_0b_cb256_steps-gamma-k} and Fig.~\ref{fig:m180_tsc_0b_cb256-gamma-k} shows that this same mode is the most unstable in both cases. However, the growth rates for TSC-640ppc are smaller due to the lower numerical noise. Note also that the growth rates for the modes M=1 and M=7 have much smaller error bars than for the run TSC-40ppc (compare with Fig.~\ref{fig:m180_tsc_0b_cb256-gamma-k}). This implies that the reduction of the numerical noise affects mainly those modes by extending the duration of their linear growth phase.
\section{Discussion and conclusions}\label{sec:conclusion}
Before drawing the conclusions from this work, it is necessary to discuss two results regarding diagnostics and simulations which are not shown here.
\subsection{Influence of the boundary conditions}
We used conducting boundary conditions in the $x$ direction in our simulations.  It has been claimed that those boundaries are not a realistic choice for applications to space plasmas. To clarify this issue, we investigated their influence on our simulation results. First, we detected waves generated at the center of the current sheet in the beginning of the simulation caused by the initial numerical noise. These waves propagate outwards becoming reflected back and forth between the conducting boundaries and the center of the current sheet. This leads leads to a periodic exchange of magnetic and kinetic energy.
The generation of such magnetosonic waves was already reported before (Refs.~\onlinecite{Ambrosiano1983,Melzani2013}). They, however, do not essentially affect the physical results. We verified this conclusion by changing the simulation box length across the reflective direction $x$ (results not shown here).
\subsection{Influence of the mass ratio}
We found that the numerical anisotropic heating is especially critical for higher mass ratios,  such as $m_i/m_e = 180$ and above, compared to the often used reduced mass ratio 25. The difference is due to the separation of the time scales of ions to electrons: $\Omega_{ci}^{-1}/\omega_{pe}^{-1}$. Indeed, while time scales of the evolution of the tearing mode instability depends only on $\Omega_{ci}^{-1}$, the number of iterations performed by the code is proportional to $\omega_{pe}^{-1}$. This has two consequences. First, the number of iterations before the onset of the same physical phenomenon is smaller for lower mass ratios (since $\Omega_{ci}^{-1}/\omega_{pe}^{-1}\propto m_i/m_e $). And second, since stochastic heating increases the kinetic energy linearly with the number of timesteps (see section 9.2 of Ref.~\onlinecite{Hockney1988}), the accumulation of numerical errors is less important for lower $m_i/m_e$.

In order to prove this claim, we ran simulations (not shown here) with the same parameters but lower mass ratio ($m_i/m_e$=25), and, thus, a smaller ratio $\Omega_{ci}^{-1}/\omega_{pe}^{-1}$.  We found no significant numerical heating with a CIC scheme.  As expected, there is no important difference between the evolution of energy and temperatures anisotropies with either TSC or CIC shape function. This is the reason why no bifurcation was seen in the past PIC-code simulations, when lower mass ratios were used. In addition to that, growth rates of the tearing mode do not show important differences between runs using different schemes. Nevertheless, it is known that the choice of too small mass ratios can reveals misleading results~\cite{Karimabadi2005a}. Therefore, the choice of higher order shape functions becomes important when more realistic results (for higher mass ratios) are desired.

The mass ratio is also a key parameter for the initial evolution in the initially anisotropic runs discussed in Sec.~\ref{sec:anisotropic_runs}. Indeed, both Weibel and mirror instabilities have growth rates that depends on the mass ratio $m_i/m_e$ (at least, in the numerical solutions for the fully kinetic expressions not shown here). By choosing more realistic values, the growth rates are in general increased. This implies that a PIC simulation with a too low mass ratio will underestimate the importance of those instabilities, leading to an incorrect physical evolution of the system. Moreover, the dependence of those kinetic instabilities (as well as many other ones) on $m_i/m_e$  is in general different. Therefore, the change of the mass ratio would may also imply a change in the dominant instability of the system, affecting dramatically the physical predictions of a PIC simulation (see Ref.~\onlinecite{Bret2009,Bret2010} for further details).
\subsection{Influence of the restriction to an antiparallel magnetic field configuration}\label{sec:guidefield}
It is important to remark that all the effects related with the anisotropy mentioned in this paper are obtained for an antiparallel magnetic field configuration, neglecting any finite (guide) magnetic field in the out-of-plane direction. Under the influence of a (strong enough) guide field, electrons become magnetized (gyrotropic) in this guide field  with a highly isotropic temperature in the reconnection plane. As a result, the stabilization and saturation of the tearing mode in the guide field case will not depend on the electron temperature anisotropy. Instead of that, it has more to do with processes like the electron trapping in magnetic islands, etc. (see, e.g., Refs.~\onlinecite{Karimabadi2005a,Karimabadi2005}).

Therefore, the proper modelling of the electron temperature anisotropy is especially critical for the antiparallel field case, while in case of a finite guide field, other quantities are more important for the stabilization of the tearing mode, such as the resolution of the electron Larmor radius on the guide field.
\subsection{Conclusions}
By means of PIC simulations of thin collisionless Harris current sheets, we have shown that higher order macroparticle shape functions allow to obtain more realistic (less numerically determined) PIC code results. In particular, the use of the TSC scheme results in a much better energy conservation over long times than the traditionally used CIC shape function. In an efficient way, the TSC shape function suppresses numerical collisions and anisotropic heating better than the CIC shape function, i.e., at less computational cost than a larger number of macroparticles per cell. This is in agreement with the results of a previous study of laser wakefield interactions by PIC code simulations\cite{Cormier-Michel2008}. We found that the choice of a TSC shape function is especially important for larger (more realistic) mass ratios and for long PIC runs.

Furthermore, lower order shape functions (such as the often used CIC) or a too small number of macroparticles per cell enhance the numerical collisions inherent to all PIC codes, leading to non-physical results. Since numerical collisions correspond to irreversible processes, they enhance the entropy of the system. We found that for higher order shape functions (TSC instead of CIC), the increase of entropy slows down more than by the use of a larger number of macroparticles per cell (see Fig.~\ref{fig:entropies}). This is because the diffusion coefficient of the effective Boltzmann equation for the macroparticles, as a measure of the collisionality, depends explicitly on the shape function (see Eqs.~\eqref{collision_operator} and \eqref{collision}).

Numerical collisions correspond to an effective scattering. They cause unphysical numerical heating\cite{Cormier-Michel2008}, leading to an artificial increase in the total energy of the system (see Fig.~\ref{fig:0b_temp_comparison}(a)). Note that the heating due to effective scattering is different from  grid heating~\cite{Langdon1970}. The latter depends mainly on the number of macroparticles per cell. It becomes dominant when the grid size is larger than the Debye length, i.e.: not in our case. Numerical heating affects mostly electrons.
This in agreement with previous simulations and analytical estimations (See section 9.2 of Ref.~\onlinecite{Hockney1988}). It was shown that the numerical heating is inversely proportional to the mass of the macroparticles considered, being thus negligible for the ions. It is also important to remark that the heating time depends strongly on the type of shape function, while the collision time does not (it only has a weakly dependence on the shape function. See Eq.~9.22 of Ref.~\onlinecite{Hockney1988} or Ref.~\onlinecite{Hockney1971} for the 2D electrostatic case).
Following Ref.~\onlinecite{Hockney1971}, the heating time measures the energy losses in an ideal homogeneous plasma without external electromagnetic fields, ideally being infinite in that system. In our case, we expect physical heating of the plasma only at later (non-linear) stages of the tearing mode evolution, due to particle acceleration caused by the reconnected electric field in the out-of-plane direction.

The observed electron
heating is anisotropic, as one can see in Fig.~\ref{fig:m180_natural anisotropies}. This was already noticed in an early work~\cite{Matsuda1975} and has to do with the anisotropic nature of collisions in 2D3V PIC code simulations in magnetic fields.

It is known that temperature anisotropies can affect the stability of a current sheet. Indeed, we confirmed the theoretical anisotropy threshold for the stabilization of the tearing mode instability\cite{Chen1984}, given by Eqs.~\eqref{anisotropy} or~\eqref{anisotropy2}, by imposing anisotropy and by letting it develop in simulations with low order shape functions or too small number of macroparticles per cell.

We also confirmed that the temperature anisotropy drives additional micro-instabilities. For strong enough (electron) temperature anisotropies, as they develop in simulations with low order shape functions (CIC), prone to numerical heating, current sheets bifurcate by a reduction of the out-of-plane component of the current density $J_z$ around their center, which corresponds to a reduction of the out-of-plane component of the electron bulk velocity $V_{e,z}$.
In agreement with previous studies, we confirmed that bifurcated current sheets inhibit the development of the tearing mode instability.  Bifurcation of Harris-type current sheets occurs around near their center, where the magnetic field vanishes. It can be due to the electron chaotic scattering~\cite{Lee2012f} after anisotropies are enhanced by, e.g., numerical collisions in electromagnetic PIC code simulations. Note that these collisions are inversely proportional to the magnetic field strength\cite{Matsuda1975}.

We proved that a Harris current sheet can spontaneously bifurcate even in schemes with good energy conservation (TSC shape function), in case of an initialization with a high enough level of electron temperature anisotropy  $A_e$. Only in the limit of negligible scattering, current sheets do not bifurcate. For our parameters, the threshold for the onset of bifurcation is around $A_e\gtrsim1.2$, just a little bit above the analytical estimate for tearing mode stabilization given by Eq.~\eqref{anisotropy2} (which is related with the Weibel instability criterion). Although this threshold was derived for a collisionless Vlasov plasma, it can still be useful for predicting bifurcation for levels of numerical collisions feasible to find in typical PIC simulations.

We also clarified the mechanism of the initial temperature relaxation anisotropy seen in some of the cases with an initially imposed electron temperature anisotropy. Although the initial state with a temperature anisotropy $A_e$ is still a Harris equilibrium, under the right conditions it can be unstable, in addition to the tearing mode, to the Weibel instability close to the unmagnetized center of the current sheet. A high enough initially imposed electron anisotropy (see threshold condition Eq.~\ref{weibel_threshold}) will necessarily have to be relaxed shortly after the initial stage via Weibel instability, at much shorter timescales than the ones associated with the slow growth of tearing mode. This instability generates filamented structures of the out-of-plane magnetic field  $B_z$, relaxing the anisotropy in addition to magnetize the center of the current sheet. In addition, this magnetization provides the conditions for the triggering of the mirror instability afterwards. Both instabilities grow together until the saturation stage, but the regions around the center affected by the mirror instability will develop a bifurcation of $J_z$.  The details of the evolution depends on the number of macroparticles per cell, since the growth of the Weibel generated magnetic field has to start from the magnetic field originated from the initial numerical noise, but this physical process should always be observed under the right conditions.

By isolating the effects of initially imposed electron or ion temperature anisotropies, we confirmed that only the electron temperature anisotropy $A_e$ is important for the stabilization of the tearing mode, even if there is an ion temperature anisotropy $A_i\neq0$: the latter does not play a role in this process, at least for the small values of $A_i$ analyzed here. An ion temperature anisotropy can stabilize a tearing mode only if $A_e$ is practically zero and $A_i$ is very large\cite{Quest2010a}. Moreover, the fast relaxation of an initially imposed temperature anisotropy via Weibel/mirror instabilities does not take place with only $A_i\neq0$. This is because the Weibel generated magnetic field that provide the conditions for the mirror modes,  and the associated bifurcation later on, requires an electron anisotropy to be triggered.

Finally, taking into account all those considerations,  we found growth rates of the simulated tearing instability that match the theoretical predictions (see Figs.~\ref{fig:m180_tsc_0b_cb256-gamma-k} and \ref{fig:m180_tsc_640ppc_0b_cb256_steps-gamma-k}). There are discrepancies, however, that are produced by the nonlinear exchange of energy among different modes allowed in a multimode tearing situation, and to a lesser degree by the enhanced noise level (coarse graining effect),  due to the reduced number of physical particles represented by one macroparticle in a PIC code.

\begin{acknowledgments}
	We acknowledge the developers of the code ACRONYM at W\"urzburg University, the support by the Max-Planck-Princeton Center for Plasma Physics and of the DFG-CRC 963 for P.K.  P.M acknowledges the financial support of the Max-Planck Society via the IMPRS for Solar System Research.

	All authors thank the referee for valuable and constructive comments which helped us to verify our results and to make our points more clear.
\end{acknowledgments}

\appendix

\section{Shape function}\label{sec:shape_function}
The use of a spatial grid in PIC codes implies that every macroparticle should be distributed in the real space, as described by a shape function $S(x)$\cite{Birdsall1991,Hockney1988}. Shape functions distribute the charge of the macroparticles over the grid and determinate the action of the electromagnetic forces on the macroparticles. $S(x)$ can be defined by the following expression for a single-macroparticle distribution function,
\begin{equation}\label{eq:shape_function}
	f(\vec{x},\vec{v})=R\,\delta(\vec{v}-\vec{v}_n) S(\vec{x} - \vec{x}_n),
\end{equation}
where $\vec{x}_n$ is the position around which the macroparticle is distributed (smeared out),  $\vec{v}_n$ is its velocity and $R$ is the ratio of the number of physical particles to (numerical) macroparticles. It is appropriate to normalize $S(x)$ to 1 and choose it separable:  $S(\vec{x} - \vec{x}_n)=S(x - x_n)S(y - y_n)S(z - z_n)$~\cite{Haugbolle2013}. Eq.~\eqref{eq:shape_function} implies that the macroparticles have a finite spatial extent. It can be characterized by the weight function $W(x)$, which is obtained by integrating the shape function over the cell volume
\begin{equation}
	W(\vec{x}_c - \vec{x}_n) =\int_{\vec{x}_c -  \vec{\Delta x}/2}^{\vec{x}_c + \vec{\Delta x}/2}  S(\vec{x}'-\vec{x}_n) d\vec{x}'.
\end{equation}
Hence, $W(x)$ characterizes the overlap of the spatially distributed macroparticle and the grid cell of size $\vec{\Delta x}$ centered around the mesh point $\vec{x}_c$.

In practice, only a small number of different shape functions are used. Higher order functions are generally more accurate (since they reduce the aliasing effects associated with the undersampling by the interpolation~\cite{Haugbolle2013}). On the other hand, they are also computationally more expensive. The use of higher order shape functions reduce the error in the calculation of non-physical forces between the macroparticles smeared over the grid. Thus, the corresponding PIC simulated plasma behaves more similar to a real plasma~\cite{Eastwood1974}.

The simplest shape function is the zero-order weighting scheme called NGP (Nearest Grid Point), with the (one-dimensional) weight function
\begin{equation}
	W(x)=\begin{cases}
	1  &  \text{ if $\xi<1/2$},\\
	0 & \text{otherwise.}
	\end{cases}
\end{equation}
Here, $\xi= |x - x_n|/\Delta x$ is the relative distance with respect to the center of the macroparticle. A NGP function weighting assigns all the charge of a macroparticle to its nearest grid point. The corresponding $S(x)$ is a Dirac delta function: each macroparticle is concentrated at one location. Although fast, the NGP scheme is rarely used since it is very noisy (due to the enhanced numerical scattering among the macroparticles).

The next higher order weighting scheme can be obtained through a convolution of the NGP weighting function with itself (the so-called b-splines are obtained by applying this method successive times. See, e.g, Ref.~\onlinecite{Cormier-Michel2008}). This CIC (Cloud in Cell) scheme, the most used one in PIC codes, is defined by the weight function:
\begin{equation}
	W(x)=\begin{cases}
	1 - \xi &  \text{ if $\xi<1$,}\\
	0 & \text{otherwise}.
	\end{cases}
\end{equation}
The CIC scheme distributes the charge of every macroparticle between the two nearest grid points by means of a linear interpolation. It provides a good compromise between the smoothness of the interparticle force and computational speed.

\begin{figure}[H]
	\centering
	\includegraphics[width=0.7\linewidth]{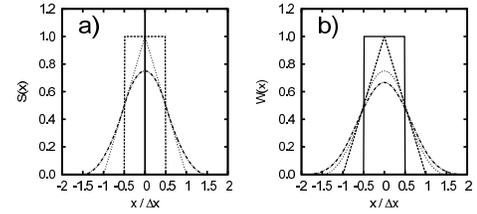}
	\caption{Left panel: 1D shape functions $S(x)$. Right panel:  1D weight functions $W(x)$. Continuous line: NGP. Dashed line: CIC. Dotted line: TSC. Dash-dotted: PQS. Those figures are for a macroparticle located at the origin $\vec{x}_n=0$. }\label{fig:1dshapes}
\end{figure}

\begin{figure}[H]
	\centering
	\includegraphics[width=0.7\linewidth]{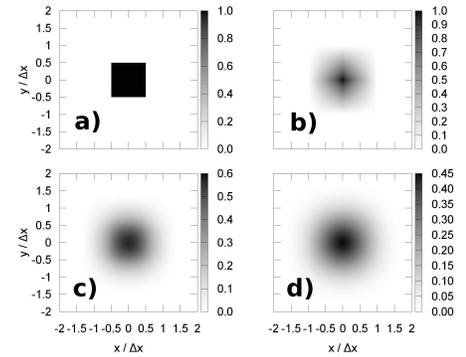}
	\caption{2D weight functions $W(x)W(y)$ for different weighting schemes. a) NGP (Nearest Grid Point). b) CIC (Cloud in Cell). c) TSC (Triangular Shaped Cloud). d) PQS (Piecewise Quadratic Spline).}\label{2dshapes}
\end{figure}

Next, there is a second order scheme called TSC (Triangular Shaped Cloud), with a quadratic weight function given by:
\begin{equation}
	W(x)=\begin{cases}
	\frac{3}{4} - \xi^2 &  \text{ if $\xi<1/2$,}\\
	\frac{1}{2}\left(\frac{3}{2} - \xi\right)^2 &  \text{ if $1/2<\xi<3/2$,}\\
	0 & \text{otherwise.}
	\end{cases}
\end{equation}
A TSC weighting scheme is much smoother than the CIC scheme. It is, however, computationally more expensive and therefore not so often used in PIC codes calculations. A third order (cubic) scheme is the PQS (Piecewise Quadratic Spline), with a weight function:
\begin{equation}
	W(x)=\begin{cases}
	\frac{1}{6}\left(4 - 6\xi^2 + 3\xi^3\right) &  \text{ if $0<\xi<1$,}\\
	\frac{1}{6}\left(2 - \xi^3\right) &  \text{ if $1<\xi<2$,}\\
	0 & \text{otherwise.}
	\end{cases}
\end{equation}
These different shape functions and their associated weight functions are illustrated in Fig.~\ref{fig:1dshapes} for 1D and in Fig. \ref{2dshapes} for 2D.
There are three consequences of choosing successive higher order shape functions. First, an increase of the width  of $W(x)$ by $\Delta x$. Second, a smoother continuity~\cite{Cormier-Michel2008} of the charge density distribution and the electromagnetic fields. Those functions increase their continuity class from $C^{n}$ to $C^{n+1}$, with $n$ the order of the interpolation scheme. The third and most important consequence for the goal of this paper is the reduction of numerical collisions between the macroparticles by using higher order shape functions~\cite{Matsuda1975}. See next Appendix~\ref{sec:collisions} for their role in PIC codes.
\section{Numerical collisions}\label{sec:collisions}
Numerical collisions are principally inherent to all PIC codes. In fact, due to the calculations on a spatial grid, a finite time step and the representation of many physical particles by macroparticles (coarse-graining), PIC codes do not model a strictly collisionless plasma,  which would fully obey the Vlasov-Maxwell equations  (see Ref.~\onlinecite{Melzani2013} for a extended discussion about this issue).
Instead of the Vlasov equation, PIC codes practically solve a kinetic Boltzmann equation for the distribution function, with an effective numerical collision operator absent in the Vlasov equation. This collision operator can be estimated as being proportional to~\cite{Birdsall1991}:
\begin{eqnarray}\label{collision}
	\left(\frac{\partial f }{\partial t}\right)_c && \propto \int \vec{dk}\frac{\vec{\kappa}\vec{\kappa}}{K^4}\frac{\tilde{S}^2(\vec{k})}{|\epsilon(\vec{k},\vec{k}\cdot\vec{v})|^2}\sum_{p=-\infty}^{\infty}\tilde{S}^2(\vec{k}_p) \\
	                                              && \times\int d\vec{v}'\delta(\vec{k}\cdot\vec{v}-\vec{k}_p\cdot\vec{v}',\omega_g).\nonumber
\end{eqnarray}
For more detailed discussion, see appendix E of Ref.~\onlinecite{Birdsall1991}. Here $\vec{\kappa}$ and $K$ are the finite difference gradient and Laplacian operators, respectively (associated with $\vec{k}$ and $k^2$, respectively). $\vec{k}_p=\vec{k}-\vec{p}\cdot \vec{k}_g $, $\vec{k}_g=2\pi/\Delta (\vec{x}^{-1})^T$ is the grid wave number, $\omega_g=2\pi/\Delta t$ is the characteristic frequency of the time stepping, $\delta(\omega,\omega_g)=\sum_{q=-\infty}^{\infty}\delta(\omega-q\omega_g)$ is a periodic delta-function comb. $\epsilon$ is the plasma dielectric function  which does not only depend on the physical model being analyzed, but also depends on the time integration scheme, the conservation properties of the algorithm, the shape function and other details of a PIC code.

As one can see in Eq.~\eqref{collision}, the collision operator involves directly the Fourier transform of the shape function $\tilde{S}(\vec{k})$ as well as the grid size $\Delta x$ and the simulation time step $\Delta t$. It can be rearranged in the form of a  Fokker-Plank term with effective diffusion $D_{ij}$ and drag $A_i$ coefficients:
\begin{equation}\label{collision_operator}
	\left(\frac{\partial f }{\partial t}\right)_c = \frac{\partial}{\partial v_i}D_{ij}\frac{\partial f}{\partial v_j} +  \frac{\partial}{\partial v_i} A_i f .
\end{equation}
The coefficients $D_{ij}$ and $A_i$  depend on the spread of the macroparticle shape function in comparison with the Debye length, and not so much on the number of macroparticles per cell. This is in contrast to real plasmas with individual point particles (see chp. 12 of Ref.~\onlinecite{Birdsall1991}, section 7.5 of Ref.~\onlinecite{Hockney1988} and Ref.~\onlinecite{Okuda1970}). Since  $\tilde{S}(\vec{k})$ decays faster for higher order shape functions, the values of the diffusion coefficients decrease as well as the influence of the numerical collisions.

Finally, it is important to remark that numerical collisions can be interpreted as a scattering of macroparticles, and therefore they can cause a non-physical particle heating (the second numerical heating mechanism discussed in Sec.~\ref{sec:intro}).

\end{document}